\documentclass[aip,reprint]{revtex4-1}

\usepackage{graphicx}
\usepackage{xcolor}

\draft

\begin{document}

\title{Stacked Josephson junctions for quantum circuit applications} 
\author{Alex Kreuzer}\affiliation{Physikalisches Institut, Karlsruhe Institute of Technology, 76131 Karlsruhe, Germany}
\author{Thilo Krumrey}\affiliation{Physikalisches Institut, Karlsruhe Institute of Technology, 76131 Karlsruhe, Germany}
\author{Hossam Tohamy}\affiliation{Physikalisches Institut, Karlsruhe Institute of Technology, 76131 Karlsruhe, Germany}
\author{Alexandru Ionita}\affiliation{Physikalisches Institut, Karlsruhe Institute of Technology, 76131 Karlsruhe, Germany}
\author{Hannes Rotzinger}\altaffiliation{corresponding author: rotzinger@kit.edu}\affiliation{Physikalisches Institut, Karlsruhe Institute of Technology, 76131 Karlsruhe, Germany}\affiliation{Institute for Quantum Materials and Technologies, Karlsruhe Institute of Technology, 76131 Karlsruhe, Germany}
\author{Alexey V. Ustinov}\affiliation{Physikalisches Institut, Karlsruhe Institute of Technology, 76131 Karlsruhe, Germany}\affiliation{Institute for Quantum Materials and Technologies, Karlsruhe Institute of Technology, 76131 Karlsruhe, Germany}

\date{\today}

\begin{abstract}
Low-loss inductors are essential components in various superconducting circuits, such as qubits or digital electronics. In this study, we investigate highly compact inductors formed by vertical stacking of Josephson junctions. Our implementation employs multiple layers of aluminum separated by tunnel barriers. Individual stacks are connected by suspended superconducting bridges, which are free of additional dielectric materials and therefore should not contribute significantly to losses. 
We present implementation details, fabrication results, and device characterization measurements.
\end{abstract}
\maketitle

The distribution of electrical currents in superconducting circuits often requires the use of inductors. This is particularly the case in quantum circuits, where dissipative circuit elements such as resistors are typically avoided due to their Nyquist contribution to the overall noise. Inductive elements are realized either by creating a magnetic field, for instance by using a sufficiently long wire, or by exploiting the kinetic energy of the charge carriers. Wires made of a material with low charge carrier density (see e.g.\cite{Dacca_1997, Mitra2016, Rotzinger2016, Niepce2019, Kirsh2021, Kristen2023, Kristen2024}) or Josephson tunnel contacts are particularly suitable as kinetic inductors because of their small size and very low stray magnetic field. Notable applications include a variety of qubit types\cite{Mooij1999, Manucharyan2009, Quarton2020}, compact microwave resonators\cite{Zhang2019, Basset2019, Frasca2023, Yang2024}, travelling wave parametric amplifiers\cite{White2015, Macklin2015, Faramarzi2024, Giachero2024} and superconducting digital electronic circuits\cite{Altimiras2013, Tolpygo2016, Tolpygo2018, Fox2019, CastellanosBeltran2019, Tolpygo2023}. 

\emph{Arrays} of Josephson junctions are advantageous when a linear current-phase relation with a high inductance is required (up to $\mu$H, see e.g. \cite{Pechenezhskiy2020,Bell2012}). The junctions are typically arranged side-by-side on a dielectric substrate with dimensions ranging from micrometers to nanometers. The advantages of using junction arrays as inductors are in their well-predictable properties and ease of fabrication.

\begin{figure}
\includegraphics[width=8.5cm, height=2.51cm]{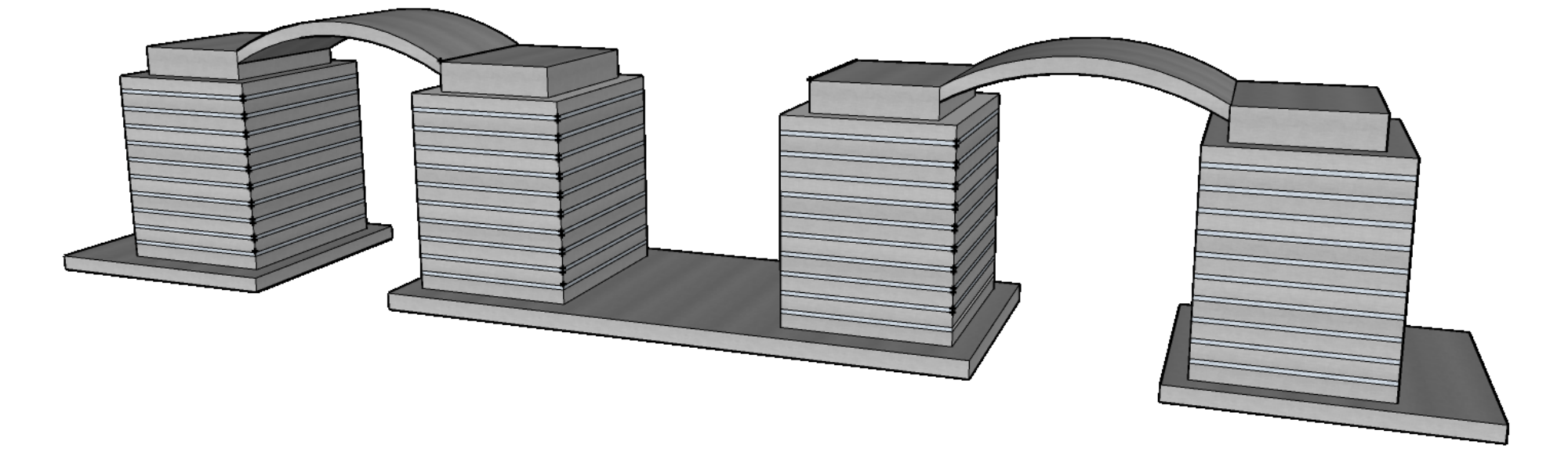}
\caption{Schematics of an array of stacked junctions with a bridge arch connecting the top of two stacks. The stacks and the arches are not to scale.\label{fig:Schematics}}
\end{figure}

In this paper, we seek to reduce the on-chip footprints of Josephson junction arrays by employing a stacked configuration with the ultimate objective of developing inductors for highly integrated superconducting circuits. The proposed approach is independent of the number of junctions per stack and allows the total inductance to be tailored by varying the number of Josephson junctions and their individual kinetic inductances. Given that the footprint area of the stack $A$ is constant, the kinetic inductance is solely dependent on the critical current density of each junction. We connect two neighboring stacks by a suspended superconducting bridge, thus adding no additional dielectric material in the vicinity of the inductor (see Fig.~\ref{fig:Schematics}). 

The study of stacked junction arrays has a long history in the field of superconducting electronics. For example, stacks have been used to model the electrodynamics of layered high-temperature superconductors\cite{Ustinov1993, Sakai1994, Pedersen1995, Ustinov1996}, to improve the properties of Josephson oscillators\cite{Shitov1996}, and are also used as lumped inductors for superconducting digital circuits \cite{Fox2019, CastellanosBeltran2019}. 

When developing a large inductive impedance $Z =\omega L> Z_0 \approx 377\,\Omega$, it is crucial to consider the stray capacitance of the individual inductive elements. The combination of self- and stray capacitances and junction inductances gives rise to plasmon-like oscillatory modes, which can pollute the frequency spectrum and degrade device performance\cite{Hutter2011,Masluk2012,Maleeva2018}.
The reduction of parasitic capacitance is based on the use of superconducting islands of the smallest feasible dimensions. In the planar configuration, this necessitates the utilization of electron-beam lithography or a comparable high-resolution lithography technique. For applications requiring extremely high impedances, an elegant approach has been developed in the past, in which the planar junction array is under-etched and separated from the dielectric substrate \cite{Pechenezhskiy2020}. In this work, we introduce junction stacking as an alternative technique to significantly reduce the stray capacitance of the array.

The paper is organized as follows. First, we evaluate the expected stray capacitance of Josephson junction stacks. After that, we describe the fabrication process and present the measurement results for two types of Josephson junction stacks.

\section{Stray capacitance estimates}
\begin{figure}
\includegraphics[width=8.5cm, height=12.55cm]{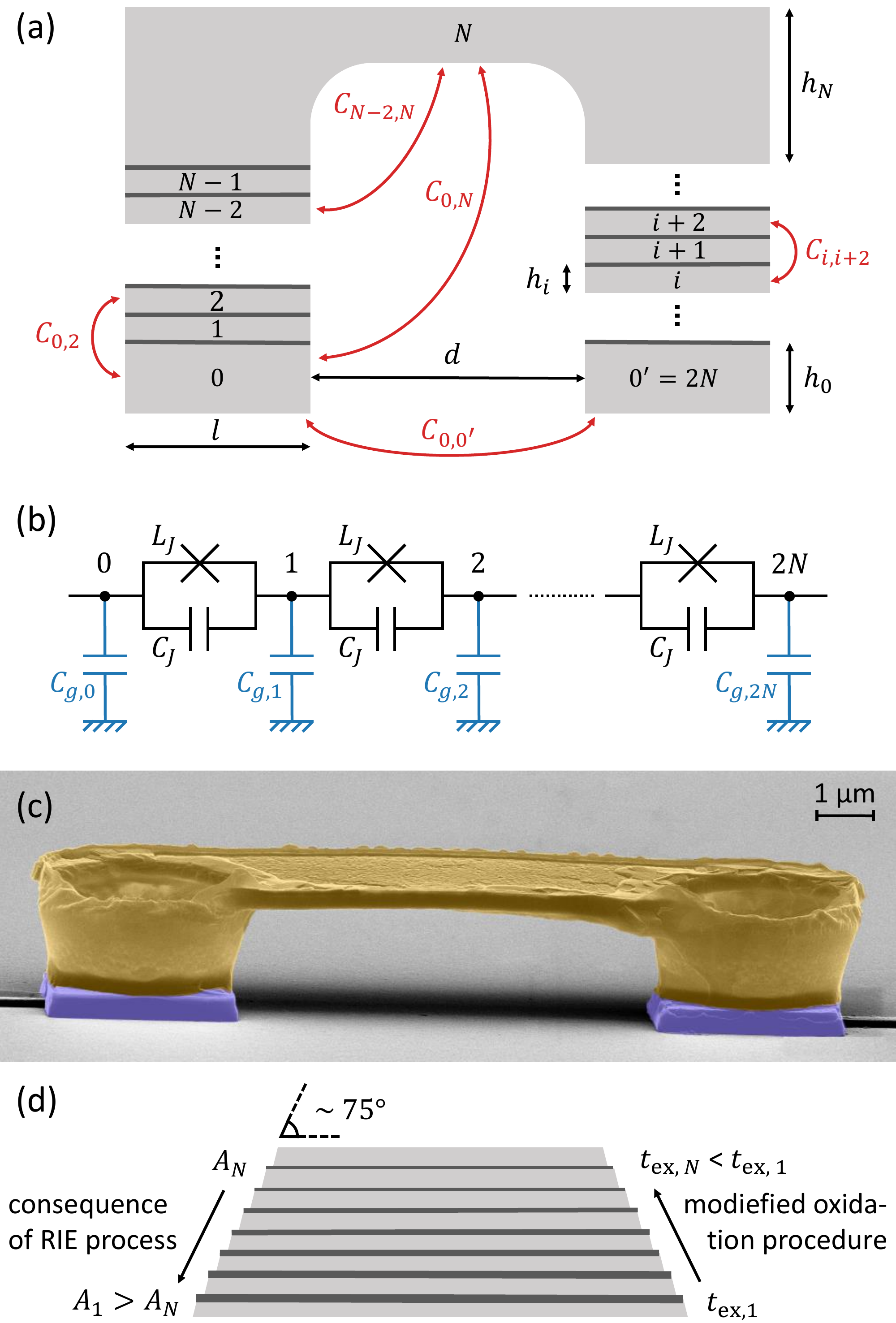}
\caption{(a) Schematics used for estimation of the stray capacitance of superconducting islands.
(b) Schematics of an array of $2N$ Josephson junctions. $C_J$ and $L_J$ denote the capacitance and inductance of one junction, $C_{g,i}$ marks the islands capacitance to ground.
(c) A false-color scanning electron microscopy image of two stacks and an aluminum bridge (gold). Each stack consists of eight Al/AlO$_x$/Al junctions (purple).
(d) Schematics of stack cross section (not to scale). The RIE milling of the conductor (Al) and tunnel barrier (AlO$_x$) results in the formation of a slanted sidewall. In order to compensate for the differing junction areas, the oxygen exposure is decreased from the bottom to the top. \label{fig:SEM}}
\end{figure}

We consider a pair of $N$-junction stacks, each with a square footprint and $N+1$ superconducting electrodes placed on a dielectric substrate. The two stacks are connected by a superconducting bridge sketched in Fig.~\ref{fig:SEM}(a). The bottom layer has a thickness $h_0$, an area $A$ and width $l$; the stacks are separated by a distance $d$. The bridge has height $h_N$ and width $w_N$ (perpendicular to the arch). The thickness of the $i$-th stacked electrode is $h_i$.

The two stacks connected by the bridge form an array of $2N+1$ superconducting islands, where each island $i$ has a mutual capacitance $C_{i,j}$ with another island $j \neq i$. For neighboring islands, this is the capacitance of the junction $C_J$. In addition, depending on the dimensions of the stacks and the arch, the superconducting island $i$ exhibits a stray capacitance $C_{\mathrm{stray},i}$ coupling it to more distant islands and the ground $C_{g,i}$ (e.g., due to wiring or ground planes). Individual contributions of stray capacitance are estimated using a numerical finite element solver\cite{ansys_maxwell} under the following assumptions.
We assume $h_i \ll l$ and consider only the stray capacitance contribution $C_{i,i+2}$ among next but one neighboring electrode (with an effective capacitor area proportional to $l\cdot h_i$). 
Since the bottom electrode ($i=0$) is placed directly on the substrate, we consider its stray capacitance $C_{0,i}$ to each stacked electrode within the stack.
The coupling between the bottom electrodes $C_{0,0'}$ of the two connected stacks is dominated by the electric field inside the substrate (e.g., silicon, sapphire), which typically has a dielectric constant significantly higher than that of vacuum. Assuming $h_i\sim 10\,$nm and $d >1\,\mu$m, we neglect the contribution of the capacitance between the inner electrodes of the adjacent stacks. The stray capacitance between the bottom electrode and the arch is denoted as $C_{0,N}$. We consider a narrow bridge arch with $w_N < d$ and assume a linear scaling of the stray capacitance with $w_N$. Since the height of the fabricated bridges $h_N$ is comparable to their width $l$, we take into account the stray capacitance between each stacked electrode and the arch $C_{i,N}$.

\begin{figure}
\includegraphics[width=8.5cm, height=12.326cm]{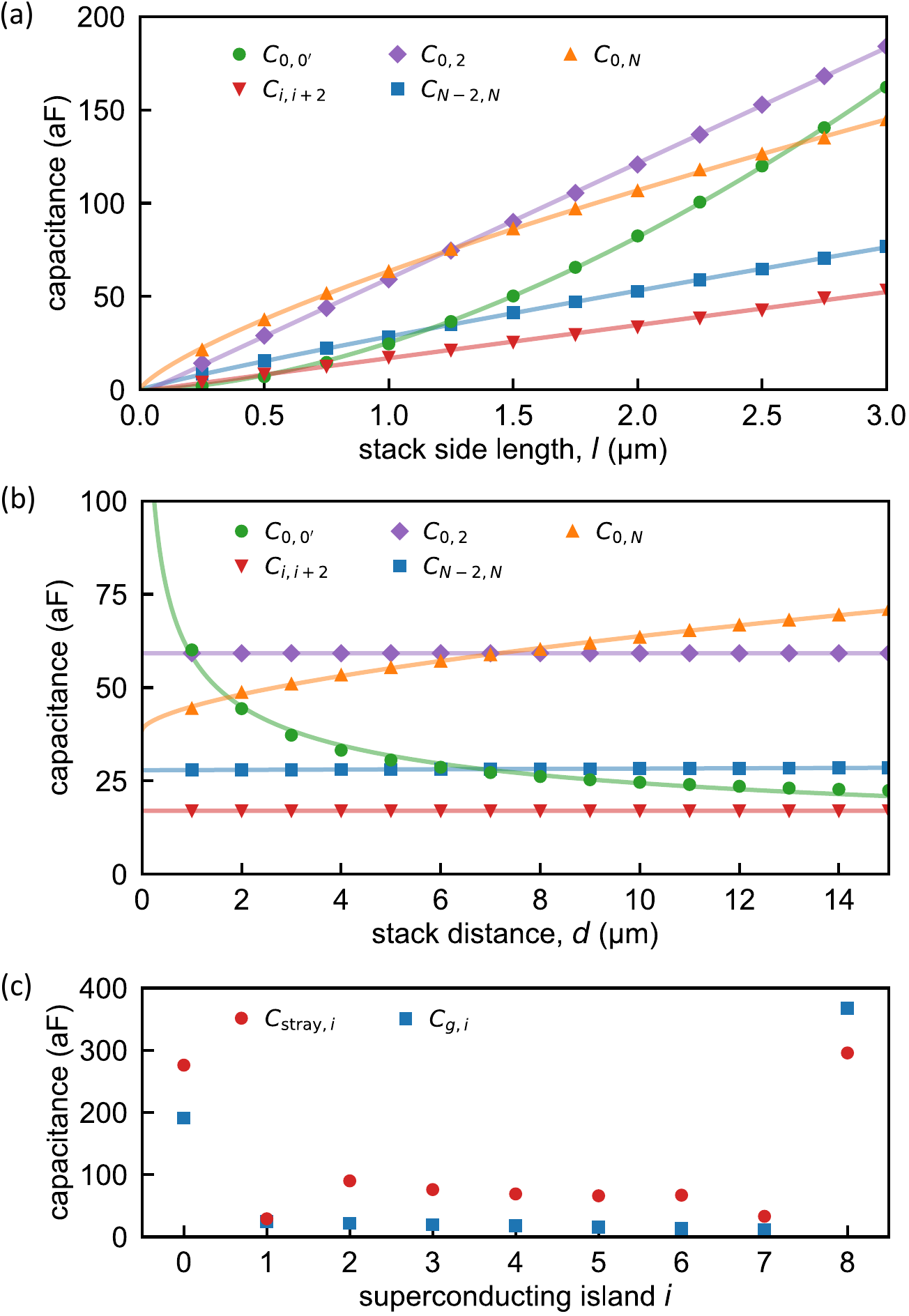}
\caption{Stray capacitance $C_{i,j}$ between pairs of superconducting islands of the junction stack. (a) Dependence on $l$ ($d$ fixed to 10$\,\mu$m), (b) Dependence on $d$ ($l$ fixed to 1$\,\mu$m). (c) Total stray capacitance $C_{\mathrm{stray},i}$ and capacitance to ground $C_{g,i}$ for each superconducting island in a stack. The islands $i=0$ and $i=8$ correspond to the bottom electrode and the bridge arch. In the three plots, the double stack arrays are modeled with the parameter $N=8$, $h_0=100\,$nm, $h_i=35\,$nm, $h_N=3\,\mu$m and $w_N=1\,\mu$m. \label{fig:C_vs_l}}
\end{figure}

The numerically estimated stray capacitance components are plotted in Fig.~\ref{fig:C_vs_l}(a) as a function of the width of the stack $l$ and in Fig.~\ref{fig:C_vs_l}(b) as a function of the distance between the two stacks $d$. 
For all stacks with $l$ below $3\,\mu$m the stray capacitance is dominated by the coupling to the bottom electrode within a stack ($C_{0,2}$ and $C_{0,N}$). For footprints larger than $1\,\mu$m the substrate contribution $C_{0,0'}$ cannot be neglected. The situation is different when considering the distance between the stacks. Here, $C_{0,0'}$ dominates for $d \leq l=1\,\mu$m, while for large $d$ the capacitance $C_{0,N}$ between the bottom electrode and the arch dominates. The numerical data are fitted to polynomial functions with parameters given in Tab.~\ref{tab:capacitances}.

Summation of all contributions 
gives the total stay capacitance of island $i$:
\begin{equation}
    C_{\mathrm{stray},i} = \sum_{j \neq i} C_{i,j}
\end{equation}
The estimated $C_{\mathrm{stray},i}$  for the stacks fabricated and measured in this paper (designed with $N=8$, $l=1\,\mu$m, $d=10\,\mu$m, $h_0=100\,$nm, $h_i=35\,$nm, $h_N=3\,\mu$m and $w_N=1\,\mu$m) is shown in Fig.~\ref{fig:C_vs_l}(c). For the stacked Al/AlO$_x$/Al junctions with a capacitance of approximately $C_J \sim 50\,$fF$/\mu$m$^2$ the estimated stray capacitance is negligible, yielding $C_{\mathrm{stray},i}/C_J \sim 10^{-3}$. Consequently, long-range mutual capacitance coupling\cite{{Haviland2000}, {Peatin2023}} can be neglected, allowing the junction stack to be modeled as a nearest-neighbor coupled junction array, as depicted in Fig.~\ref{fig:SEM}(b). 

The dispersion relation of the collective mode in a uniform junction array is given by
\begin{equation} 
\omega_k = \omega_0 \sqrt{\frac{1-\cos(\frac{\pi k}{N_J})}{\frac{C_g}{2C_J} + 1-\cos(\frac{\pi k}{N_J})}} 
\end{equation}
for $k=1,2,...,N_J-1$, where $\omega_0$ is the plasma frequency of a single junction and $N_J$ is the total number of junctions in the array.\cite{Masluk2012} The collective mode frequency is suppressed by the island's capacitance to ground $C_g$, in particular through the Coulomb screening of the Cooper pairs across the array, where $\lambda=\sqrt{C_J/C_g}$ determines the screening length.\cite{Feldman2024} The elementary block of our array contains eight stacked junctions with $C_J = 50\,$fF and $L_J = 1\,$nH. With the parameters given above we estimated $C_g$ of each superconducting island, the results are shown in Figure~\ref{fig:C_vs_l}(c). Assuming an average ground capacitance of $59\,$aF, which corresponds to $C_J/C_g \approx 850$, we expect $\omega_1 \approx 0.98 \cdot \omega_0 \approx 22.2\,$GHz. This frequency is well above the operation frequencies of typical quantum circuit applications, e.g., superconducting qubits \cite{Mooij1999, Manucharyan2009, Quarton2020}. For comparison, in the previously studied planar array of $200$ junctions\cite{Weil2015} the estimated ratio was $C_J/C_g \approx 35$. Masluk et al.\cite{Masluk2012}  investigated a planar junction array specially designed to reduce stray capacitance to the ground, achieving a ratio of $C_J/C_g \approx 1000$. They reported $\omega_1 = 0.78 \cdot \omega_0 = 14.2\,$GHz for an array with $80$ junctions placed directly on the substrate.

As the number of junctions in a stack increases, additional islands are positioned further away from the substrate, resulting in a decrease in $C_{g}$. In the case of a large array (where $N\gg1$), the stray capacitance contributions of the bottom electrode and the arch become negligible. For example, with $40$ stacked junctions, we estimate the capacitance to the ground $C_g\approx 15\,$aF. Consequently, for a double stack array with $80$ junctions, we expect the lowest mode frequency to be approximately $0.91 \cdot \omega_0 \approx 20.6\,$GHz.

\begin{table}
\caption{\label{tab:capacitances} Numerical results of the stray capacitance between parts of the junction array as a function of the stack side length $l$ and the stack distance $d$ (fits to the data shown in Fig.~\ref{fig:C_vs_l}).
}
\begin{ruledtabular}
\begin{tabular}{lccr}
  &  $l$ / $\mu$m  & $d$ / $\mu$m \\
$C_{0,0'}$ / aF & $25\cdot l^{1.7}$ & $58\cdot d^{-0.4}$ \\
$C_{0,2}$ / aF & $62\cdot l-2$ & $59$  \\
$C_{0,N}$ / aF & $64\cdot l^{0.8}$ & $6\cdot d^{0.6}+39$ \\
$C_{i,i+2}$ / aF & $18\cdot l-1$ & $17$  \\
$C_{N-2,N}$ / aF & $29\cdot l^{0.9}$ & $28$ \\
\end{tabular}
\end{ruledtabular}
\end{table}

\section{Fabrication}
The fabrication process begins with the deposition of an Al/(AlO$_x$/Al)$_N$ multilayer on a sapphire substrate. Each aluminum layer has a thickness of 35\,nm, followed by a oxygen exposure to form the tunnel barrier. 
In the next step, the stacked junctions are defined by a square resist mask and etched by a dry ArCl$_2$/ArCl$_2$O$_2$ RIE process using an ICP tool. The etching is calibrated to stop within the bottom layer, which has a thickness of 100\,nm. This layer is used for wiring and additional structures and will be defined in a later step. 
Finally, a bridge arch is deposited on the junction stacks using optical lithography, aluminum deposition, and ArCl$_2$ dry etching.

The fabrication result of an example array with $2\times 8$ junctions is shown in Fig.~\ref{fig:SEM}(c). Here, the two square junction stacks have a footprint of approximately 3\,$\mu$m × 3\,$\mu$m, while the bridge arch has a typical height of approximately 3\,$\mu$m and a length of 10\,$\mu$m.

\begin{figure}
\includegraphics[width=8.5cm, height=7.8cm]{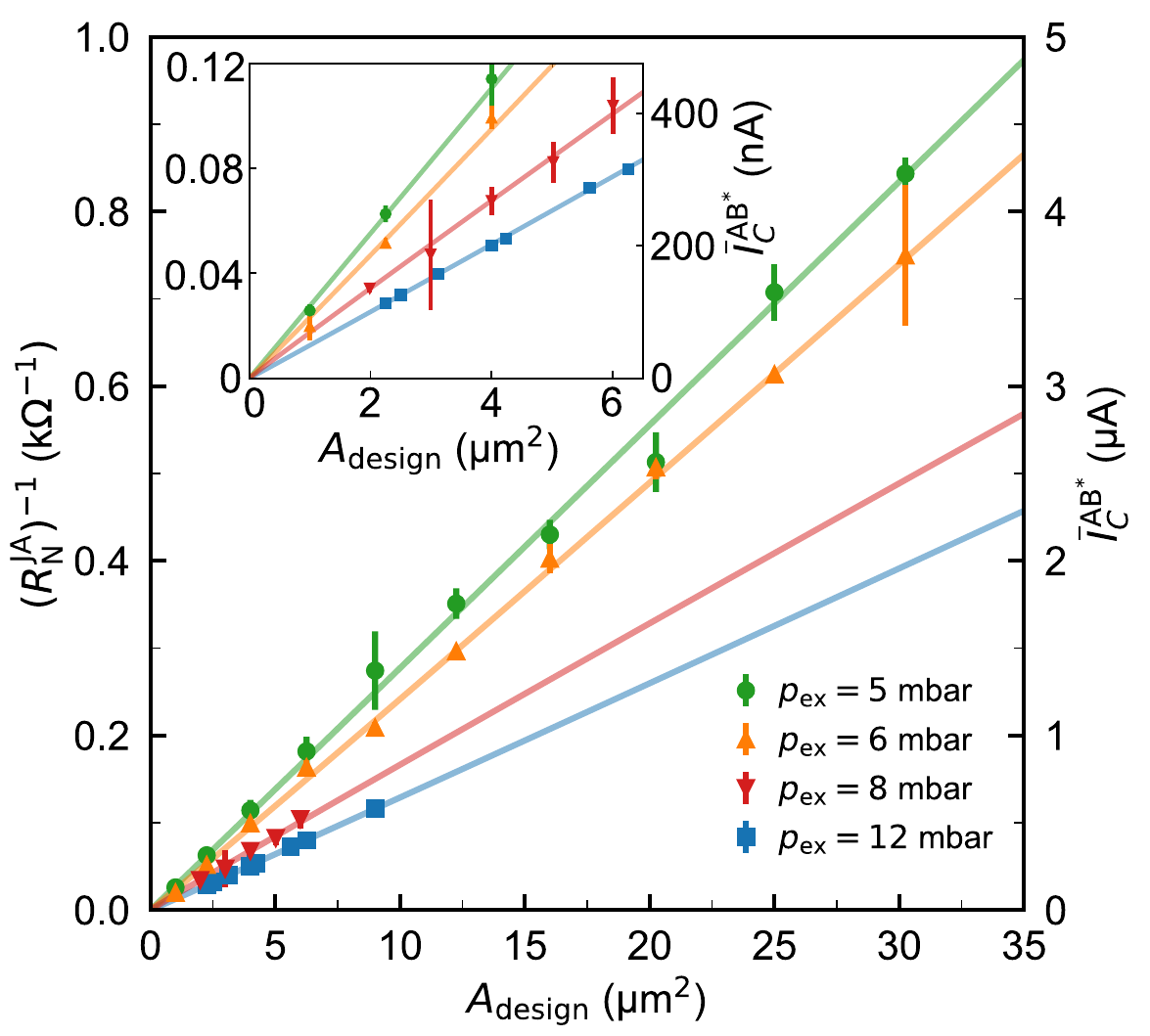}
\caption{Dependence of the inverse normal resistance $(R_{\mathrm{N}}^{\mathrm{JA}})^{-1}$ of arrays comprising $2\times8=16$ junctions on the designed junction area $A_{\mathrm{design}}$ for four different wafers and varied junction's oxygen exposure pressure. The y-axis on the right side gives the estimate of the derived average critical current $\overline{I}_C^{\mathrm{AB^*}}$ per junction within the array. \label{fig:Rn_vs_A}}
\end{figure}

To study the dependence of the tunnel barrier on oxygen exposure, we fabricated four wafers with different base pressures and systematically varied the junction size between $1~$µm$^2$ and $30~$µm$^2$. 
Figure~\ref{fig:Rn_vs_A} shows the median values of 20 individual test structures each. The data is consistent with the expected behavior of $A_{\mathrm{design}} \propto 1/R_{\mathrm{N}}^{\mathrm{JA}}$ (solid line). The extracted specific junction resistances are listed in the table \ref{tab:Rn_fit}.

\begin{table}
\caption{\label{tab:Rn_fit} Specific junction resistance at a given tunnel barrier oxidation pressure (same data as in Fig.~\ref{fig:Rn_vs_A}).
}
\begin{ruledtabular}
\begin{tabular}{lccccr}
$p_{\mathrm{ex}}$ / mbar  &  5    & 6    & 8    & 12 \\
$\rho$ / k$\Omega/$µm$^2$ &  36.3 & 43.2 & 57.5 & 79.6 \\
\end{tabular}
\end{ruledtabular}
\end{table}

\begin{figure}
\includegraphics[width=8.5cm, height=7.665cm]{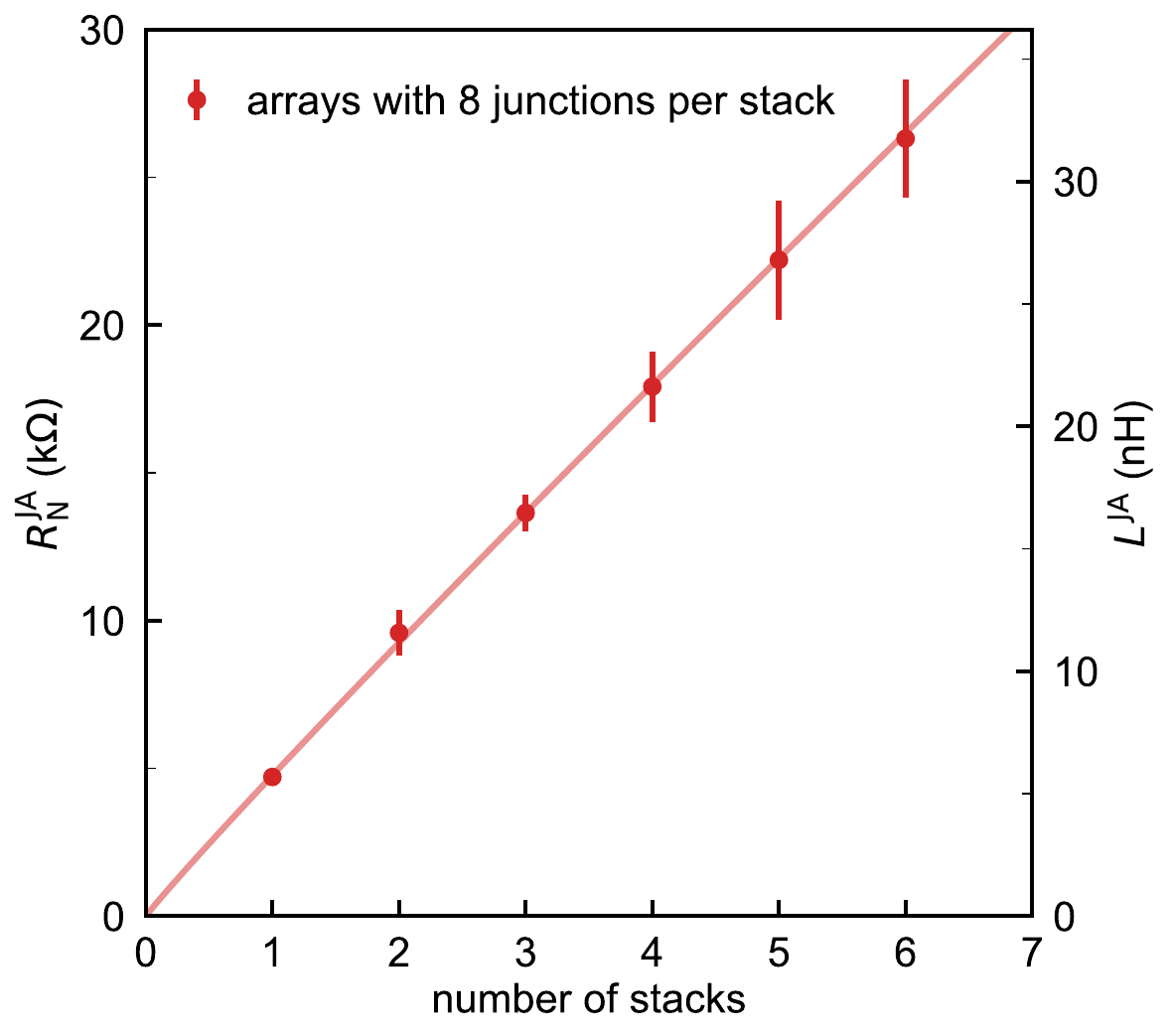}
\caption{Dependence of the normal resistance $R_{\mathrm{N}}^{\mathrm{JA}}$ on the number of stacks ($n$) in arrays. Each stack consists of eight junctions.  The error bars indicate the standard deviation over 20 measured arrays. The solid line is a fit to  $R_{\mathrm{N}}^{\mathrm{JA}}= \rho n ^b$ and is in good agreement with $b=1$. The second y-axis shows an estimated for the expected inductance.\label{fig:Rn_vs_n}}
\end{figure}

In a separate experiment, we have fabricated arrays of $8$ junctions per stack. Here, the footprint ($2 \times 2\,\mu$m$^2$) and a oxygen barrier pressure ($6\,$mbar) is kept fixed but the total number of stacks varied.

Figure \ref{fig:Rn_vs_n} gives the median values of $R^{\mathrm{JA}}_{\mathrm{N}}$ of 20 samples for each array. The data were compared to $R_{\mathrm{N}}^{\mathrm{JA}} = \rho {n}^b$, where $n$ is the number of stacks and agrees well with $b=0.96$ and $\rho = 4.8~$k$\Omega$. From the well known Ambegaokar-Baratoff\cite{Ambegaokar1963} relation, we can estimate the critical current, and thus the expected specific inductance ($\sim5.4~$nH per stack).

\section{Uncompensated Josephson junction arrays}
\begin{figure}
\includegraphics[width=8.5cm, height=8.5cm]{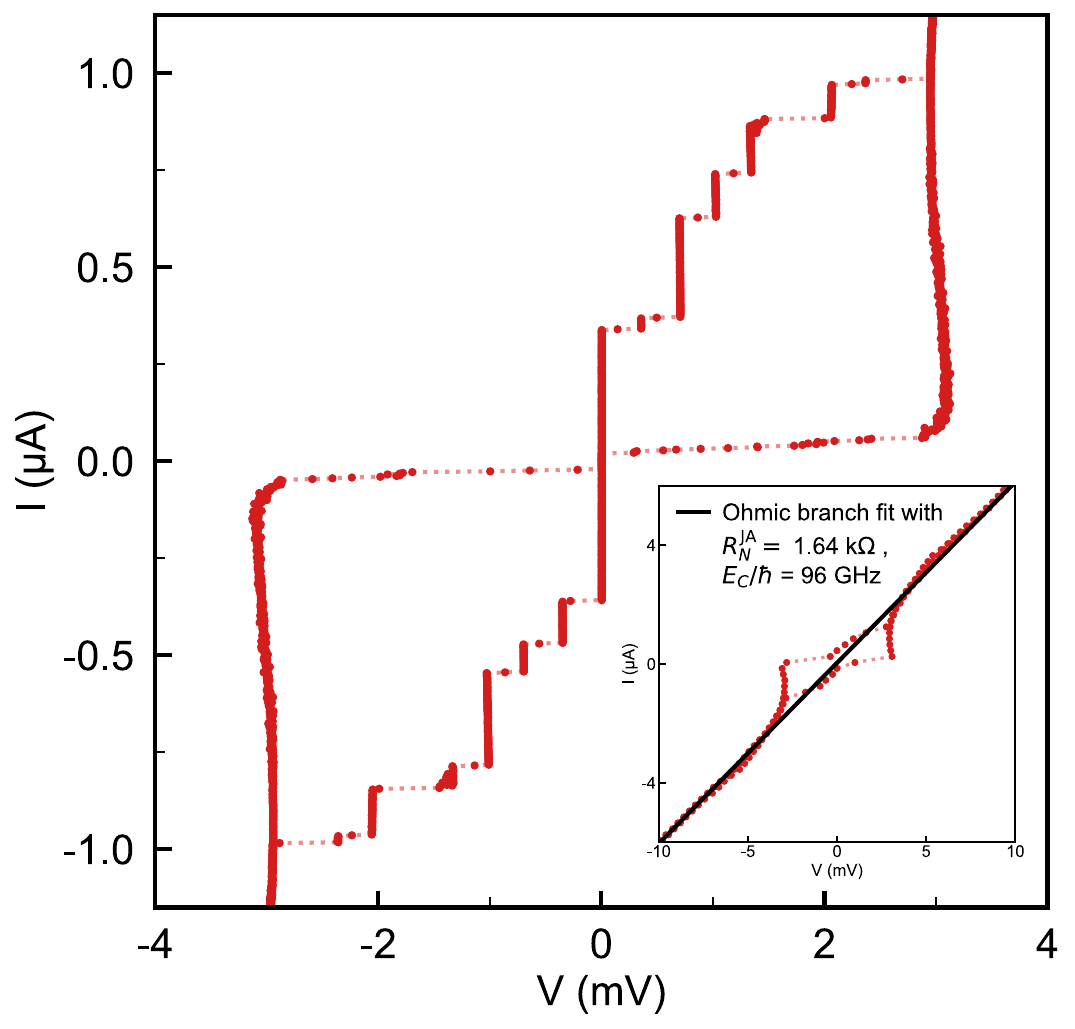}
\caption{IVC of an uncompensated array consisting of $2\times4=8$ stacked junctions with a footprint of $4\times4\,\mu$m$^2$. The individual junctions in the array switch to the normal state at bias currents between 350\,nA and 1\,$\mu$A, see text for details.\label{fig:SteppedHysteresis}}
\end{figure}

A typical current-voltage characteristic (IVC) of a sample compromising eight junctions is shown in Fig.~\ref{fig:SteppedHysteresis} with a footprint of $4\times4~$µm$^2$. 
The IVC exhibits hysteresis typical of underdamped Josephson junctions with a McCumber parameter of $\beta_C = (4I_C/\pi I_r)^2 \approx 10^3$. We measure a ratio of sub-gap resistance to normal resistance, $R_{\mathrm{sg}}/ R_{\mathrm{N}}^{\mathrm{JA}}$ of about 40. From the total gap voltage of 2.84\,mV we estimate an average single junction gap voltage of $\bar{V}_{\mathrm{gap}} \approx 355\,\mu$V. This is close to the expected BCS value $2\Delta /e = 360$\,$\mu$V of Al at low temperatures, see e.g.\cite{Lotkhov2006}. A comparison of $V=R_{\mathrm{N}}^{\mathrm{JA}} I + E_C / e$, to the Ohmic branch, yields a normal resistance $R_{\mathrm{N}}^{\mathrm{JA}}$ of $1.64~$k$\Omega$ and a charging energy $E_C/\hbar$ of $96~$GHz for the array.

An interesting feature is shown on the critical current branch. Here, the IVC shows six voltage discontinuities that are multiples of $\bar{V}_{\mathrm{gap}}$. We interpret these as reduced $I_C$ of individual junction in the array that switch to the normal conducting state at different bias current values.
Since all junctions in the fabrication should nominally have the same current density, we attribute the distribution of $I_C$ to a varying area of the individual junctions. This assumption is supported by SEM images (not shown).

For most applications it is desirable to have junctions in the array with a uniform $I_C$ distribution, ideally with a constant $I_C$. Then, due to the constant inductance fraction, the phase $\varphi$ would drop equally across each junction. For small phase drops with a sufficiently large number of junctions ($\Delta \varphi < \pi/4$), the inductance contribution of each individual junction is approximately linear and the total array inductance scales with the number of junctions\cite{{Basko2013},{Nguyen2017}}. This is not the case if the critical current spread is large, since then the individual contribution of junctions with a smaller $I_C$ can lead to significant non-linearity.

\section{Compensated Josephson junction arrays}
\begin{figure}
\includegraphics[width=8.5cm, height=8.5cm]{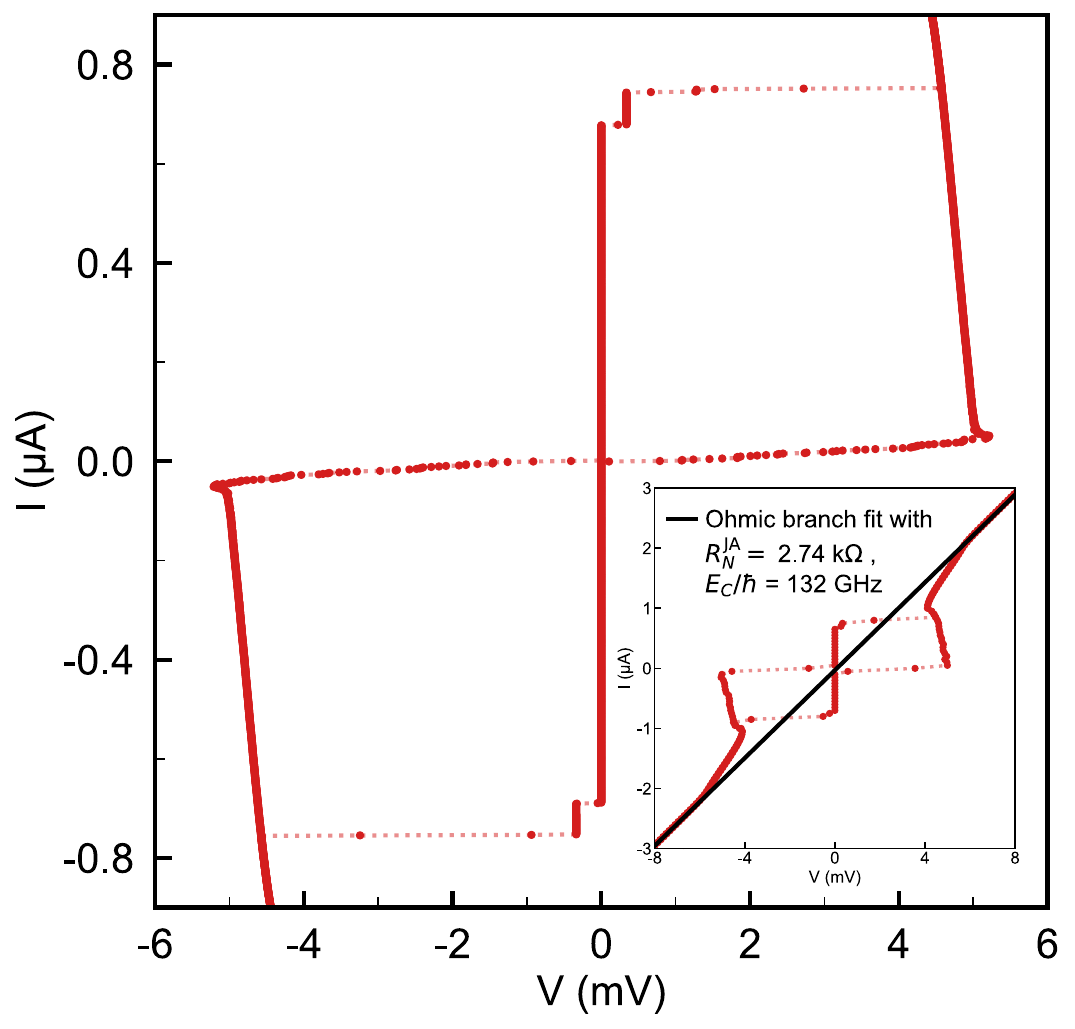}
\caption{IV-curve of an compensated array compromising $2\times8=16$ stacked junctions with a footprint of $3\times3~$µm$^2$. The stacked junctions were oxidized with increasing exposure times to take care of the systematic change in junction area (compare with figure \ref{fig:SteppedHysteresis}). Typical junction's $I_C$ are close to $750~$nA. The McCumber parameter $\beta_C \sim 10^5$ and the sub-gap resistance to the normal resistance ratio $R_{\mathrm{sg}}/R_{\mathrm{N}}^{\mathrm{JA}} \sim 10^2$ with $R_{\mathrm{sg}}~100~$k$\Omega$ are obtained from the measured data. \label{fig:Hysteresis}}
\end{figure}

We identify the RIE etch step as the origin of the slanted etch profile, mainly because the chemistry used etches aluminum much more efficiently than AlO$_2$. This effectively leads to a staircase like structure, visualized in Fig.~\ref{fig:SEM}(d). A consequence of the side wall profile is an increase of the junction areas from top to bottom.

We tested two strategies to mitigate the effect of the $I_C$ distribution in the stacks. First, we optimize the etch recipe to increase the side slope from $50^\circ$ to $75^\circ$. This increases the uniformity of the individual junctions, but also allows  for stacks with a much smaller footprint. 

The second strategy relies on a fundamental difference between planar and stacked arrays. In most planar configurations, each Josephson junction is defined with the same tunnel barrier parameters, i.e., dielectric material and thickness. Because the tunnel barriers are subsequently fabricated in a stacked geometry, the individual barrier parameters or even the material in each layer can be tweaked. 
In particular, we have modified the oxidation procedure in the fabrication process to account for the different junction sizes by gradually reducing the oxidation times from the bottom to the top of the stack. The exposure of the $ith$ stacked junction was adjusted using the well-known empirical trend for the tunnel barrier resistance as a function of oxygen exposure pressure $\times$ time, as outlined in\cite{Kleinsasser1995}. We aim for an approximately equal $R$ for all junctions $(E_i/E_1)^{\alpha} \approx A_i/A_1$, where $i=1$ refers to the bottom junction and area $A_1$ of the stack. We use an empirically determined exponent $\alpha \approx 0.4$ for the exposure range of $10^2$ to $10^4~$mbar$\cdot$s.

In the deposition process, we keep the oxygen exposure pressure constant and vary the individual junction oxidation times. They are set based on the ratio of the individual area to the stack footprint, $t_{\mathrm{ex},i} = t_{\mathrm{ex},1} (A_i/A_1)^{1/\alpha}$. 

Figure \ref{fig:Hysteresis} shows the current-voltage characteristics of an array consisting of $16$ junctions with a footprint of $3\times3~$µm$^2$, fabricated using the compensation technique. 
The junctions were oxidized with decreasing exposure times from $12$ to $6$ minutes at a pressure of 5\,mbar.  

While most of the IVC are very similar, the step like structure of uncompensated arrays is not visible anymore. We also note a $\beta_C \sim 10^5$ and the sub-gab resistance on the order of $100~$k$\Omega$ ($R_{\mathrm{sg}}/R_{\mathrm{N}}^{\mathrm{JA}} \sim 10^2$).
From the Ambegaokar-Baratoff relation, we expect a critical current of 1.6\,$\mu$A, which is about a factor of two larger than the switching current observed in the IVC. We attribute this difference to a noise or phase diffusion induced suppression. Assuming an uniform $I_C$ distribution, we expect an inductance of $3.3~$nH for the junction array.

\section{Conclusion}
We have demonstrated that stacks of Josephson junctions can be fabricated to form compact and high-quality inductors with low stray capacitance and McCumber parameters on the order of $10^5$. The arrays can contain an almost arbitrary number of junctions and do not require additional dielectric layers, making them attractive for superconducting quantum circuit applications. Area variations in the stack are compensated by fine-tuning the tunnel barrier, resulting in increased uniformity of the critical current.

\section{Acknowledgment}
We acknowledge the support of Silvia Diewald and Aina Quintilla from the KIT Nanostructure Service Laboratory. This research was supported by the German Federal Ministry of Education and Research under the Research Program Quantum Systems, through the projects GeQCoS (FZK13N15691) and qBriqs (FZK13N15950).

\section*{Data Availability}
The data presented in this study are available upon reasonable request.

\section*{References}
\bibliography{References.bib}

\begin{thebibliography}{44}%
\makeatletter
\providecommand \@ifxundefined [1]{%
 \@ifx{#1\undefined}
}%
\providecommand \@ifnum [1]{%
 \ifnum #1\expandafter \@firstoftwo
 \else \expandafter \@secondoftwo
 \fi
}%
\providecommand \@ifx [1]{%
 \ifx #1\expandafter \@firstoftwo
 \else \expandafter \@secondoftwo
 \fi
}%
\providecommand \natexlab [1]{#1}%
\providecommand \enquote  [1]{``#1''}%
\providecommand \bibnamefont  [1]{#1}%
\providecommand \bibfnamefont [1]{#1}%
\providecommand \citenamefont [1]{#1}%
\providecommand \href@noop [0]{\@secondoftwo}%
\providecommand \href [0]{\begingroup \@sanitize@url \@href}%
\providecommand \@href[1]{\@@startlink{#1}\@@href}%
\providecommand \@@href[1]{\endgroup#1\@@endlink}%
\providecommand \@sanitize@url [0]{\catcode `\\12\catcode `\$12\catcode `\&12\catcode `\#12\catcode `\^12\catcode `\_12\catcode `\%12\relax}%
\providecommand \@@startlink[1]{}%
\providecommand \@@endlink[0]{}%
\providecommand \url  [0]{\begingroup\@sanitize@url \@url }%
\providecommand \@url [1]{\endgroup\@href {#1}{\urlprefix }}%
\providecommand \urlprefix  [0]{URL }%
\providecommand \Eprint [0]{\href }%
\providecommand \doibase [0]{http://dx.doi.org/}%
\providecommand \selectlanguage [0]{\@gobble}%
\providecommand \bibinfo  [0]{\@secondoftwo}%
\providecommand \bibfield  [0]{\@secondoftwo}%
\providecommand \translation [1]{[#1]}%
\providecommand \BibitemOpen [0]{}%
\providecommand \bibitemStop [0]{}%
\providecommand \bibitemNoStop [0]{.\EOS\space}%
\providecommand \EOS [0]{\spacefactor3000\relax}%
\providecommand \BibitemShut  [1]{\csname bibitem#1\endcsname}%
\let\auto@bib@innerbib\@empty
%</preamble>
\bibitem [{\citenamefont {Dacca}\ \emph {et~al.}(1997)\citenamefont {Dacca}, \citenamefont {Gemme}, \citenamefont {Musenich}, \citenamefont {Parodi}, \citenamefont {Pittaluga}, \citenamefont {Rizzini},\ and\ \citenamefont {Buscaglia}}]{Dacca_1997}%
  \BibitemOpen
  \bibfield  {author} {\bibinfo {author} {\bibfnamefont {A.}~\bibnamefont {Dacca}}, \bibinfo {author} {\bibfnamefont {G.}~\bibnamefont {Gemme}}, \bibinfo {author} {\bibfnamefont {R.}~\bibnamefont {Musenich}}, \bibinfo {author} {\bibfnamefont {R.}~\bibnamefont {Parodi}}, \bibinfo {author} {\bibfnamefont {S.}~\bibnamefont {Pittaluga}}, \bibinfo {author} {\bibfnamefont {S.}~\bibnamefont {Rizzini}}, \ and\ \bibinfo {author} {\bibfnamefont {V.}~\bibnamefont {Buscaglia}},\ }\bibfield  {title} {\enquote {\bibinfo {title} {{Niobium titanium nitride for superconducting accelerating cavities}},}\ }in\ \href@noop {} {\emph {\bibinfo {booktitle} {{8th Workshop on RF Superconductivity}}}}\ (\bibinfo {year} {1997})\ pp.\ \bibinfo {pages} {1103--1109}\BibitemShut {NoStop}%
\bibitem [{\citenamefont {Mitra}\ \emph {et~al.}(2016)\citenamefont {Mitra}, \citenamefont {Tewari}, \citenamefont {Mahalu},\ and\ \citenamefont {Shahar}}]{Mitra2016}%
  \BibitemOpen
  \bibfield  {author} {\bibinfo {author} {\bibfnamefont {S.}~\bibnamefont {Mitra}}, \bibinfo {author} {\bibfnamefont {G.~C.}\ \bibnamefont {Tewari}}, \bibinfo {author} {\bibfnamefont {D.}~\bibnamefont {Mahalu}}, \ and\ \bibinfo {author} {\bibfnamefont {D.}~\bibnamefont {Shahar}},\ }\bibfield  {title} {\enquote {\bibinfo {title} {Negative magnetoresistance in amorphous indium oxide wires},}\ }\href {\doibase 10.1038/srep37687} {\bibfield  {journal} {\bibinfo  {journal} {Scientific Reports}\ }\textbf {\bibinfo {volume} {6}} (\bibinfo {year} {2016}),\ 10.1038/srep37687}\BibitemShut {NoStop}%
\bibitem [{\citenamefont {Rotzinger}\ \emph {et~al.}(2016)\citenamefont {Rotzinger}, \citenamefont {Skacel}, \citenamefont {Pfirrmann}, \citenamefont {Voss}, \citenamefont {M{\"u}nzberg}, \citenamefont {Probst}, \citenamefont {Bushev}, \citenamefont {Weides}, \citenamefont {Ustinov},\ and\ \citenamefont {Mooij}}]{Rotzinger2016}%
  \BibitemOpen
  \bibfield  {author} {\bibinfo {author} {\bibfnamefont {H.}~\bibnamefont {Rotzinger}}, \bibinfo {author} {\bibfnamefont {S.}~\bibnamefont {Skacel}}, \bibinfo {author} {\bibfnamefont {M.}~\bibnamefont {Pfirrmann}}, \bibinfo {author} {\bibfnamefont {J.}~\bibnamefont {Voss}}, \bibinfo {author} {\bibfnamefont {J.}~\bibnamefont {M{\"u}nzberg}}, \bibinfo {author} {\bibfnamefont {S.}~\bibnamefont {Probst}}, \bibinfo {author} {\bibfnamefont {P.}~\bibnamefont {Bushev}}, \bibinfo {author} {\bibfnamefont {M.}~\bibnamefont {Weides}}, \bibinfo {author} {\bibfnamefont {A.}~\bibnamefont {Ustinov}}, \ and\ \bibinfo {author} {\bibfnamefont {J.}~\bibnamefont {Mooij}},\ }\bibfield  {title} {\enquote {\bibinfo {title} {Aluminium-oxide wires for superconducting high kinetic inductance circuits},}\ }\href@noop {} {\bibfield  {journal} {\bibinfo  {journal} {Superconductor Science and Technology}\ }\textbf {\bibinfo {volume} {30}},\ \bibinfo {pages} {025002} (\bibinfo {year} {2016})}\BibitemShut {NoStop}%
\bibitem [{\citenamefont {Niepce}, \citenamefont {Burnett},\ and\ \citenamefont {Bylander}(2019)}]{Niepce2019}%
  \BibitemOpen
  \bibfield  {author} {\bibinfo {author} {\bibfnamefont {D.}~\bibnamefont {Niepce}}, \bibinfo {author} {\bibfnamefont {J.}~\bibnamefont {Burnett}}, \ and\ \bibinfo {author} {\bibfnamefont {J.}~\bibnamefont {Bylander}},\ }\bibfield  {title} {\enquote {\bibinfo {title} {High kinetic inductance nbn nanowire superinductors},}\ }\href {\doibase 10.1103/physrevapplied.11.044014} {\bibfield  {journal} {\bibinfo  {journal} {Physical Review Applied}\ }\textbf {\bibinfo {volume} {11}} (\bibinfo {year} {2019}),\ 10.1103/physrevapplied.11.044014}\BibitemShut {NoStop}%
\bibitem [{\citenamefont {Kirsh}\ \emph {et~al.}(2021)\citenamefont {Kirsh}, \citenamefont {Svetitsky}, \citenamefont {Goldstein}, \citenamefont {Pardo}, \citenamefont {Hachmo},\ and\ \citenamefont {Katz}}]{Kirsh2021}%
  \BibitemOpen
  \bibfield  {author} {\bibinfo {author} {\bibfnamefont {N.}~\bibnamefont {Kirsh}}, \bibinfo {author} {\bibfnamefont {E.}~\bibnamefont {Svetitsky}}, \bibinfo {author} {\bibfnamefont {S.}~\bibnamefont {Goldstein}}, \bibinfo {author} {\bibfnamefont {G.}~\bibnamefont {Pardo}}, \bibinfo {author} {\bibfnamefont {O.}~\bibnamefont {Hachmo}}, \ and\ \bibinfo {author} {\bibfnamefont {N.}~\bibnamefont {Katz}},\ }\bibfield  {title} {\enquote {\bibinfo {title} {Linear and nonlinear properties of a compact high-kinetic-inductance wsi multimode resonator},}\ }\href {\doibase 10.1103/physrevapplied.16.044017} {\bibfield  {journal} {\bibinfo  {journal} {Physical Review Applied}\ }\textbf {\bibinfo {volume} {16}} (\bibinfo {year} {2021}),\ 10.1103/physrevapplied.16.044017}\BibitemShut {NoStop}%
\bibitem [{\citenamefont {Kristen}\ \emph {et~al.}(2023)\citenamefont {Kristen}, \citenamefont {Voss}, \citenamefont {Wildermuth}, \citenamefont {Rotzinger},\ and\ \citenamefont {Ustinov}}]{Kristen2023}%
  \BibitemOpen
  \bibfield  {author} {\bibinfo {author} {\bibfnamefont {M.}~\bibnamefont {Kristen}}, \bibinfo {author} {\bibfnamefont {J.~N.}\ \bibnamefont {Voss}}, \bibinfo {author} {\bibfnamefont {M.}~\bibnamefont {Wildermuth}}, \bibinfo {author} {\bibfnamefont {H.}~\bibnamefont {Rotzinger}}, \ and\ \bibinfo {author} {\bibfnamefont {A.~V.}\ \bibnamefont {Ustinov}},\ }\bibfield  {title} {\enquote {\bibinfo {title} {Random telegraph fluctuations in granular microwave resonators},}\ }\href {\doibase 10.1063/5.0147430} {\bibfield  {journal} {\bibinfo  {journal} {Applied Physics Letters}\ }\textbf {\bibinfo {volume} {122}} (\bibinfo {year} {2023}),\ 10.1063/5.0147430}\BibitemShut {NoStop}%
\bibitem [{\citenamefont {Kristen}\ \emph {et~al.}(2024)\citenamefont {Kristen}, \citenamefont {Voss}, \citenamefont {Wildermuth}, \citenamefont {Bilmes}, \citenamefont {Lisenfeld}, \citenamefont {Rotzinger},\ and\ \citenamefont {Ustinov}}]{Kristen2024}%
  \BibitemOpen
  \bibfield  {author} {\bibinfo {author} {\bibfnamefont {M.}~\bibnamefont {Kristen}}, \bibinfo {author} {\bibfnamefont {J.~N.}\ \bibnamefont {Voss}}, \bibinfo {author} {\bibfnamefont {M.}~\bibnamefont {Wildermuth}}, \bibinfo {author} {\bibfnamefont {A.}~\bibnamefont {Bilmes}}, \bibinfo {author} {\bibfnamefont {J.}~\bibnamefont {Lisenfeld}}, \bibinfo {author} {\bibfnamefont {H.}~\bibnamefont {Rotzinger}}, \ and\ \bibinfo {author} {\bibfnamefont {A.~V.}\ \bibnamefont {Ustinov}},\ }\bibfield  {title} {\enquote {\bibinfo {title} {Giant two-level systems in a granular superconductor},}\ }\href {\doibase 10.1103/PhysRevLett.132.217002} {\bibfield  {journal} {\bibinfo  {journal} {Phys. Rev. Lett.}\ }\textbf {\bibinfo {volume} {132}},\ \bibinfo {pages} {217002} (\bibinfo {year} {2024})}\BibitemShut {NoStop}%
\bibitem [{\citenamefont {Mooij}\ \emph {et~al.}(1999)\citenamefont {Mooij}, \citenamefont {Orlando}, \citenamefont {Levitov}, \citenamefont {Tian}, \citenamefont {van~der Wal},\ and\ \citenamefont {Lloyd}}]{Mooij1999}%
  \BibitemOpen
  \bibfield  {author} {\bibinfo {author} {\bibfnamefont {J.~E.}\ \bibnamefont {Mooij}}, \bibinfo {author} {\bibfnamefont {T.~P.}\ \bibnamefont {Orlando}}, \bibinfo {author} {\bibfnamefont {L.}~\bibnamefont {Levitov}}, \bibinfo {author} {\bibfnamefont {L.}~\bibnamefont {Tian}}, \bibinfo {author} {\bibfnamefont {C.~H.}\ \bibnamefont {van~der Wal}}, \ and\ \bibinfo {author} {\bibfnamefont {S.}~\bibnamefont {Lloyd}},\ }\bibfield  {title} {\enquote {\bibinfo {title} {Josephson persistent-current qubit},}\ }\href {\doibase 10.1126/science.285.5430.1036} {\bibfield  {journal} {\bibinfo  {journal} {Science}\ }\textbf {\bibinfo {volume} {285}},\ \bibinfo {pages} {1036–1039} (\bibinfo {year} {1999})}\BibitemShut {NoStop}%
\bibitem [{\citenamefont {Manucharyan}\ \emph {et~al.}(2009)\citenamefont {Manucharyan}, \citenamefont {Koch}, \citenamefont {Glazman},\ and\ \citenamefont {Devoret}}]{Manucharyan2009}%
  \BibitemOpen
  \bibfield  {author} {\bibinfo {author} {\bibfnamefont {V.~E.}\ \bibnamefont {Manucharyan}}, \bibinfo {author} {\bibfnamefont {J.}~\bibnamefont {Koch}}, \bibinfo {author} {\bibfnamefont {L.~I.}\ \bibnamefont {Glazman}}, \ and\ \bibinfo {author} {\bibfnamefont {M.~H.}\ \bibnamefont {Devoret}},\ }\bibfield  {title} {\enquote {\bibinfo {title} {Fluxonium: Single cooper-pair circuit free of charge offsets},}\ }\href {\doibase 10.1126/science.1175552} {\bibfield  {journal} {\bibinfo  {journal} {Science}\ }\textbf {\bibinfo {volume} {326}},\ \bibinfo {pages} {113–116} (\bibinfo {year} {2009})}\BibitemShut {NoStop}%
\bibitem [{\citenamefont {Yan}\ \emph {et~al.}(2020)\citenamefont {Yan}, \citenamefont {Sung}, \citenamefont {Krantz}, \citenamefont {Kamal}, \citenamefont {Kim}, \citenamefont {Yoder}, \citenamefont {Orlando}, \citenamefont {Gustavsson},\ and\ \citenamefont {Oliver}}]{Quarton2020}%
  \BibitemOpen
  \bibfield  {author} {\bibinfo {author} {\bibfnamefont {F.}~\bibnamefont {Yan}}, \bibinfo {author} {\bibfnamefont {Y.}~\bibnamefont {Sung}}, \bibinfo {author} {\bibfnamefont {P.}~\bibnamefont {Krantz}}, \bibinfo {author} {\bibfnamefont {A.}~\bibnamefont {Kamal}}, \bibinfo {author} {\bibfnamefont {D.~K.}\ \bibnamefont {Kim}}, \bibinfo {author} {\bibfnamefont {J.~L.}\ \bibnamefont {Yoder}}, \bibinfo {author} {\bibfnamefont {T.~P.}\ \bibnamefont {Orlando}}, \bibinfo {author} {\bibfnamefont {S.}~\bibnamefont {Gustavsson}}, \ and\ \bibinfo {author} {\bibfnamefont {W.~D.}\ \bibnamefont {Oliver}},\ }\bibfield  {title} {\enquote {\bibinfo {title} {Engineering framework for optimizing superconducting qubit designs},}\ }\href {https://arxiv.org/abs/2006.04130} {\bibfield  {journal} {\bibinfo  {journal} {arXiv:2006.04130}\ } (\bibinfo {year} {2020})}\BibitemShut {NoStop}%
\bibitem [{\citenamefont {Zhang}\ \emph {et~al.}(2019)\citenamefont {Zhang}, \citenamefont {Kalashnikov}, \citenamefont {Lu}, \citenamefont {Kamenov}, \citenamefont {DiNapoli},\ and\ \citenamefont {Gershenson}}]{Zhang2019}%
  \BibitemOpen
  \bibfield  {author} {\bibinfo {author} {\bibfnamefont {W.}~\bibnamefont {Zhang}}, \bibinfo {author} {\bibfnamefont {K.}~\bibnamefont {Kalashnikov}}, \bibinfo {author} {\bibfnamefont {W.-S.}\ \bibnamefont {Lu}}, \bibinfo {author} {\bibfnamefont {P.}~\bibnamefont {Kamenov}}, \bibinfo {author} {\bibfnamefont {T.}~\bibnamefont {DiNapoli}}, \ and\ \bibinfo {author} {\bibfnamefont {M.}~\bibnamefont {Gershenson}},\ }\bibfield  {title} {\enquote {\bibinfo {title} {Microresonators fabricated from high-kinetic-inductance aluminum films},}\ }\href {\doibase 10.1103/physrevapplied.11.011003} {\bibfield  {journal} {\bibinfo  {journal} {Physical Review Applied}\ }\textbf {\bibinfo {volume} {11}} (\bibinfo {year} {2019}),\ 10.1103/physrevapplied.11.011003}\BibitemShut {NoStop}%
\bibitem [{\citenamefont {Basset}\ \emph {et~al.}(2019)\citenamefont {Basset}, \citenamefont {Watfa}, \citenamefont {Aiello}, \citenamefont {Féchant}, \citenamefont {Morvan}, \citenamefont {Estève}, \citenamefont {Gabelli}, \citenamefont {Aprili}, \citenamefont {Weil}, \citenamefont {Kasumov}, \citenamefont {Bouchiat},\ and\ \citenamefont {Deblock}}]{Basset2019}%
  \BibitemOpen
  \bibfield  {author} {\bibinfo {author} {\bibfnamefont {J.}~\bibnamefont {Basset}}, \bibinfo {author} {\bibfnamefont {D.}~\bibnamefont {Watfa}}, \bibinfo {author} {\bibfnamefont {G.}~\bibnamefont {Aiello}}, \bibinfo {author} {\bibfnamefont {M.}~\bibnamefont {Féchant}}, \bibinfo {author} {\bibfnamefont {A.}~\bibnamefont {Morvan}}, \bibinfo {author} {\bibfnamefont {J.}~\bibnamefont {Estève}}, \bibinfo {author} {\bibfnamefont {J.}~\bibnamefont {Gabelli}}, \bibinfo {author} {\bibfnamefont {M.}~\bibnamefont {Aprili}}, \bibinfo {author} {\bibfnamefont {R.}~\bibnamefont {Weil}}, \bibinfo {author} {\bibfnamefont {A.}~\bibnamefont {Kasumov}}, \bibinfo {author} {\bibfnamefont {H.}~\bibnamefont {Bouchiat}}, \ and\ \bibinfo {author} {\bibfnamefont {R.}~\bibnamefont {Deblock}},\ }\bibfield  {title} {\enquote {\bibinfo {title} {High kinetic inductance microwave resonators made by he-beam assisted deposition of tungsten nanowires},}\ }\href {\doibase 10.1063/1.5080925} {\bibfield  {journal} {\bibinfo  {journal} {Applied
  Physics Letters}\ }\textbf {\bibinfo {volume} {114}} (\bibinfo {year} {2019}),\ 10.1063/1.5080925}\BibitemShut {NoStop}%
\bibitem [{\citenamefont {Frasca}\ \emph {et~al.}(2023)\citenamefont {Frasca}, \citenamefont {Arabadzhiev}, \citenamefont {de~Puechredon}, \citenamefont {Oppliger}, \citenamefont {Jouanny}, \citenamefont {Musio}, \citenamefont {Scigliuzzo}, \citenamefont {Minganti}, \citenamefont {Scarlino},\ and\ \citenamefont {Charbon}}]{Frasca2023}%
  \BibitemOpen
  \bibfield  {author} {\bibinfo {author} {\bibfnamefont {S.}~\bibnamefont {Frasca}}, \bibinfo {author} {\bibfnamefont {I.}~\bibnamefont {Arabadzhiev}}, \bibinfo {author} {\bibfnamefont {S.~B.}\ \bibnamefont {de~Puechredon}}, \bibinfo {author} {\bibfnamefont {F.}~\bibnamefont {Oppliger}}, \bibinfo {author} {\bibfnamefont {V.}~\bibnamefont {Jouanny}}, \bibinfo {author} {\bibfnamefont {R.}~\bibnamefont {Musio}}, \bibinfo {author} {\bibfnamefont {M.}~\bibnamefont {Scigliuzzo}}, \bibinfo {author} {\bibfnamefont {F.}~\bibnamefont {Minganti}}, \bibinfo {author} {\bibfnamefont {P.}~\bibnamefont {Scarlino}}, \ and\ \bibinfo {author} {\bibfnamefont {E.}~\bibnamefont {Charbon}},\ }\bibfield  {title} {\enquote {\bibinfo {title} {Nbn films with high kinetic inductance for high-quality compact superconducting resonators},}\ }\href {\doibase 10.1103/physrevapplied.20.044021} {\bibfield  {journal} {\bibinfo  {journal} {Physical Review Applied}\ }\textbf {\bibinfo {volume} {20}} (\bibinfo {year} {2023}),\
  10.1103/physrevapplied.20.044021}\BibitemShut {NoStop}%
\bibitem [{\citenamefont {Yang}\ \emph {et~al.}(2024)\citenamefont {Yang}, \citenamefont {He}, \citenamefont {Gao}, \citenamefont {Chen}, \citenamefont {Wu}, \citenamefont {Wang}, \citenamefont {Mu}, \citenamefont {Peng},\ and\ \citenamefont {Lin}}]{Yang2024}%
  \BibitemOpen
  \bibfield  {author} {\bibinfo {author} {\bibfnamefont {M.}~\bibnamefont {Yang}}, \bibinfo {author} {\bibfnamefont {X.}~\bibnamefont {He}}, \bibinfo {author} {\bibfnamefont {W.}~\bibnamefont {Gao}}, \bibinfo {author} {\bibfnamefont {J.}~\bibnamefont {Chen}}, \bibinfo {author} {\bibfnamefont {Y.}~\bibnamefont {Wu}}, \bibinfo {author} {\bibfnamefont {X.}~\bibnamefont {Wang}}, \bibinfo {author} {\bibfnamefont {G.}~\bibnamefont {Mu}}, \bibinfo {author} {\bibfnamefont {W.}~\bibnamefont {Peng}}, \ and\ \bibinfo {author} {\bibfnamefont {Z.}~\bibnamefont {Lin}},\ }\bibfield  {title} {\enquote {\bibinfo {title} {Kinetic inductance compact resonator with nbtin micronwires},}\ }\href {\doibase 10.1063/5.0220296} {\bibfield  {journal} {\bibinfo  {journal} {AIP Advances}\ }\textbf {\bibinfo {volume} {14}} (\bibinfo {year} {2024}),\ 10.1063/5.0220296}\BibitemShut {NoStop}%
\bibitem [{\citenamefont {White}\ \emph {et~al.}(2015)\citenamefont {White}, \citenamefont {Mutus}, \citenamefont {Hoi}, \citenamefont {Barends}, \citenamefont {Campbell}, \citenamefont {Chen}, \citenamefont {Chen}, \citenamefont {Chiaro}, \citenamefont {Dunsworth}, \citenamefont {Jeffrey}, \citenamefont {Kelly}, \citenamefont {Megrant}, \citenamefont {Neill}, \citenamefont {O'Malley}, \citenamefont {Roushan}, \citenamefont {Sank}, \citenamefont {Vainsencher}, \citenamefont {Wenner}, \citenamefont {Chaudhuri}, \citenamefont {Gao},\ and\ \citenamefont {Martinis}}]{White2015}%
  \BibitemOpen
  \bibfield  {author} {\bibinfo {author} {\bibfnamefont {T.~C.}\ \bibnamefont {White}}, \bibinfo {author} {\bibfnamefont {J.~Y.}\ \bibnamefont {Mutus}}, \bibinfo {author} {\bibfnamefont {I.~C.}\ \bibnamefont {Hoi}}, \bibinfo {author} {\bibfnamefont {R.}~\bibnamefont {Barends}}, \bibinfo {author} {\bibfnamefont {B.}~\bibnamefont {Campbell}}, \bibinfo {author} {\bibfnamefont {Y.}~\bibnamefont {Chen}}, \bibinfo {author} {\bibfnamefont {Z.}~\bibnamefont {Chen}}, \bibinfo {author} {\bibfnamefont {B.}~\bibnamefont {Chiaro}}, \bibinfo {author} {\bibfnamefont {A.}~\bibnamefont {Dunsworth}}, \bibinfo {author} {\bibfnamefont {E.}~\bibnamefont {Jeffrey}}, \bibinfo {author} {\bibfnamefont {J.}~\bibnamefont {Kelly}}, \bibinfo {author} {\bibfnamefont {A.}~\bibnamefont {Megrant}}, \bibinfo {author} {\bibfnamefont {C.}~\bibnamefont {Neill}}, \bibinfo {author} {\bibfnamefont {P.~J.~J.}\ \bibnamefont {O'Malley}}, \bibinfo {author} {\bibfnamefont {P.}~\bibnamefont {Roushan}}, \bibinfo {author} {\bibfnamefont {D.}~\bibnamefont
  {Sank}}, \bibinfo {author} {\bibfnamefont {A.}~\bibnamefont {Vainsencher}}, \bibinfo {author} {\bibfnamefont {J.}~\bibnamefont {Wenner}}, \bibinfo {author} {\bibfnamefont {S.}~\bibnamefont {Chaudhuri}}, \bibinfo {author} {\bibfnamefont {J.}~\bibnamefont {Gao}}, \ and\ \bibinfo {author} {\bibfnamefont {J.~M.}\ \bibnamefont {Martinis}},\ }\href {\doibase 10.48550/ARXIV.1503.04364} {\enquote {\bibinfo {title} {Traveling wave parametric amplifier with josephson junctions using minimal resonator phase matching},}\ } (\bibinfo {year} {2015})\BibitemShut {NoStop}%
\bibitem [{\citenamefont {Macklin}\ \emph {et~al.}(2015)\citenamefont {Macklin}, \citenamefont {O’Brien}, \citenamefont {Hover}, \citenamefont {Schwartz}, \citenamefont {Bolkhovsky}, \citenamefont {Zhang}, \citenamefont {Oliver},\ and\ \citenamefont {Siddiqi}}]{Macklin2015}%
  \BibitemOpen
  \bibfield  {author} {\bibinfo {author} {\bibfnamefont {C.}~\bibnamefont {Macklin}}, \bibinfo {author} {\bibfnamefont {K.}~\bibnamefont {O’Brien}}, \bibinfo {author} {\bibfnamefont {D.}~\bibnamefont {Hover}}, \bibinfo {author} {\bibfnamefont {M.~E.}\ \bibnamefont {Schwartz}}, \bibinfo {author} {\bibfnamefont {V.}~\bibnamefont {Bolkhovsky}}, \bibinfo {author} {\bibfnamefont {X.}~\bibnamefont {Zhang}}, \bibinfo {author} {\bibfnamefont {W.~D.}\ \bibnamefont {Oliver}}, \ and\ \bibinfo {author} {\bibfnamefont {I.}~\bibnamefont {Siddiqi}},\ }\bibfield  {title} {\enquote {\bibinfo {title} {A near–quantum-limited josephson traveling-wave parametric amplifier},}\ }\href {\doibase 10.1126/science.aaa8525} {\bibfield  {journal} {\bibinfo  {journal} {Science}\ }\textbf {\bibinfo {volume} {350}},\ \bibinfo {pages} {307–310} (\bibinfo {year} {2015})}\BibitemShut {NoStop}%
\bibitem [{\citenamefont {Faramarzi}\ \emph {et~al.}(2024)\citenamefont {Faramarzi}, \citenamefont {Stephenson}, \citenamefont {Sypkens}, \citenamefont {Eom}, \citenamefont {LeDuc},\ and\ \citenamefont {Day}}]{Faramarzi2024}%
  \BibitemOpen
  \bibfield  {author} {\bibinfo {author} {\bibfnamefont {F.}~\bibnamefont {Faramarzi}}, \bibinfo {author} {\bibfnamefont {R.}~\bibnamefont {Stephenson}}, \bibinfo {author} {\bibfnamefont {S.}~\bibnamefont {Sypkens}}, \bibinfo {author} {\bibfnamefont {B.~H.}\ \bibnamefont {Eom}}, \bibinfo {author} {\bibfnamefont {H.}~\bibnamefont {LeDuc}}, \ and\ \bibinfo {author} {\bibfnamefont {P.}~\bibnamefont {Day}},\ }\bibfield  {title} {\enquote {\bibinfo {title} {A 4–8 ghz kinetic inductance traveling-wave parametric amplifier using four-wave mixing with near quantum-limited noise performance},}\ }\href {\doibase 10.1063/5.0208110} {\bibfield  {journal} {\bibinfo  {journal} {APL Quantum}\ }\textbf {\bibinfo {volume} {1}} (\bibinfo {year} {2024}),\ 10.1063/5.0208110}\BibitemShut {NoStop}%
\bibitem [{\citenamefont {Giachero}\ \emph {et~al.}(2024)\citenamefont {Giachero}, \citenamefont {Vissers}, \citenamefont {Wheeler}, \citenamefont {Howe}, \citenamefont {Gao}, \citenamefont {Austermann}, \citenamefont {Hubmayr}, \citenamefont {Nucciotti},\ and\ \citenamefont {Ullom}}]{Giachero2024}%
  \BibitemOpen
  \bibfield  {author} {\bibinfo {author} {\bibfnamefont {A.}~\bibnamefont {Giachero}}, \bibinfo {author} {\bibfnamefont {M.}~\bibnamefont {Vissers}}, \bibinfo {author} {\bibfnamefont {J.}~\bibnamefont {Wheeler}}, \bibinfo {author} {\bibfnamefont {L.}~\bibnamefont {Howe}}, \bibinfo {author} {\bibfnamefont {J.}~\bibnamefont {Gao}}, \bibinfo {author} {\bibfnamefont {J.}~\bibnamefont {Austermann}}, \bibinfo {author} {\bibfnamefont {J.}~\bibnamefont {Hubmayr}}, \bibinfo {author} {\bibfnamefont {A.}~\bibnamefont {Nucciotti}}, \ and\ \bibinfo {author} {\bibfnamefont {J.}~\bibnamefont {Ullom}},\ }\bibfield  {title} {\enquote {\bibinfo {title} {Kinetic inductance traveling wave amplifier designs for practical microwave readout applications},}\ }\href {\doibase 10.1007/s10909-024-03078-1} {\bibfield  {journal} {\bibinfo  {journal} {Journal of Low Temperature Physics}\ }\textbf {\bibinfo {volume} {215}},\ \bibinfo {pages} {152–160} (\bibinfo {year} {2024})}\BibitemShut {NoStop}%
\bibitem [{\citenamefont {Altimiras}\ \emph {et~al.}(2013)\citenamefont {Altimiras}, \citenamefont {Parlavecchio}, \citenamefont {Joyez}, \citenamefont {Vion}, \citenamefont {Roche}, \citenamefont {Esteve},\ and\ \citenamefont {Portier}}]{Altimiras2013}%
  \BibitemOpen
  \bibfield  {author} {\bibinfo {author} {\bibfnamefont {C.}~\bibnamefont {Altimiras}}, \bibinfo {author} {\bibfnamefont {O.}~\bibnamefont {Parlavecchio}}, \bibinfo {author} {\bibfnamefont {P.}~\bibnamefont {Joyez}}, \bibinfo {author} {\bibfnamefont {D.}~\bibnamefont {Vion}}, \bibinfo {author} {\bibfnamefont {P.}~\bibnamefont {Roche}}, \bibinfo {author} {\bibfnamefont {D.}~\bibnamefont {Esteve}}, \ and\ \bibinfo {author} {\bibfnamefont {F.}~\bibnamefont {Portier}},\ }\bibfield  {title} {\enquote {\bibinfo {title} {Tunable microwave impedance matching to a high impedance source using a josephson metamaterial},}\ }\href {\doibase 10.1063/1.4832074} {\bibfield  {journal} {\bibinfo  {journal} {Applied Physics Letters}\ }\textbf {\bibinfo {volume} {103}} (\bibinfo {year} {2013}),\ 10.1063/1.4832074}\BibitemShut {NoStop}%
\bibitem [{\citenamefont {Tolpygo}(2016)}]{Tolpygo2016}%
  \BibitemOpen
  \bibfield  {author} {\bibinfo {author} {\bibfnamefont {S.~K.}\ \bibnamefont {Tolpygo}},\ }\bibfield  {title} {\enquote {\bibinfo {title} {Superconductor digital electronics: Scalability and energy efficiency issues (review article)},}\ }\href {\doibase 10.1063/1.4948618} {\bibfield  {journal} {\bibinfo  {journal} {Low Temperature Physics}\ }\textbf {\bibinfo {volume} {42}},\ \bibinfo {pages} {361–379} (\bibinfo {year} {2016})}\BibitemShut {NoStop}%
\bibitem [{\citenamefont {Tolpygo}\ \emph {et~al.}(2018)\citenamefont {Tolpygo}, \citenamefont {Bolkhovsky}, \citenamefont {Oates}, \citenamefont {Rastogi}, \citenamefont {Zarr}, \citenamefont {Day}, \citenamefont {Weir}, \citenamefont {Wynn},\ and\ \citenamefont {Johnson}}]{Tolpygo2018}%
  \BibitemOpen
  \bibfield  {author} {\bibinfo {author} {\bibfnamefont {S.~K.}\ \bibnamefont {Tolpygo}}, \bibinfo {author} {\bibfnamefont {V.}~\bibnamefont {Bolkhovsky}}, \bibinfo {author} {\bibfnamefont {D.~E.}\ \bibnamefont {Oates}}, \bibinfo {author} {\bibfnamefont {R.}~\bibnamefont {Rastogi}}, \bibinfo {author} {\bibfnamefont {S.}~\bibnamefont {Zarr}}, \bibinfo {author} {\bibfnamefont {A.~L.}\ \bibnamefont {Day}}, \bibinfo {author} {\bibfnamefont {T.~J.}\ \bibnamefont {Weir}}, \bibinfo {author} {\bibfnamefont {A.}~\bibnamefont {Wynn}}, \ and\ \bibinfo {author} {\bibfnamefont {L.~M.}\ \bibnamefont {Johnson}},\ }\bibfield  {title} {\enquote {\bibinfo {title} {Superconductor electronics fabrication process with monx kinetic inductors and self-shunted josephson junctions},}\ }\href {\doibase 10.1109/tasc.2018.2809442} {\bibfield  {journal} {\bibinfo  {journal} {IEEE Transactions on Applied Superconductivity}\ }\textbf {\bibinfo {volume} {28}},\ \bibinfo {pages} {1100212} (\bibinfo {year} {2018})}\BibitemShut {NoStop}%
\bibitem [{\citenamefont {Fox}\ \emph {et~al.}(2019)\citenamefont {Fox}, \citenamefont {Butler}, \citenamefont {Thompson}, \citenamefont {Dresselhaus},\ and\ \citenamefont {Benz}}]{Fox2019}%
  \BibitemOpen
  \bibfield  {author} {\bibinfo {author} {\bibfnamefont {A.~E.}\ \bibnamefont {Fox}}, \bibinfo {author} {\bibfnamefont {G.}~\bibnamefont {Butler}}, \bibinfo {author} {\bibfnamefont {M.}~\bibnamefont {Thompson}}, \bibinfo {author} {\bibfnamefont {P.~D.}\ \bibnamefont {Dresselhaus}}, \ and\ \bibinfo {author} {\bibfnamefont {S.~P.}\ \bibnamefont {Benz}},\ }\bibfield  {title} {\enquote {\bibinfo {title} {Induced current effects in josephson voltage standard circuits},}\ }\href {\doibase 10.1109/tasc.2019.2901886} {\bibfield  {journal} {\bibinfo  {journal} {IEEE Transactions on Applied Superconductivity}\ }\textbf {\bibinfo {volume} {29}},\ \bibinfo {pages} {1101808} (\bibinfo {year} {2019})}\BibitemShut {NoStop}%
\bibitem [{\citenamefont {Castellanos-Beltran}\ \emph {et~al.}(2019)\citenamefont {Castellanos-Beltran}, \citenamefont {Olaya}, \citenamefont {Sirois}, \citenamefont {Dresselhaus}, \citenamefont {Benz},\ and\ \citenamefont {Hopkins}}]{CastellanosBeltran2019}%
  \BibitemOpen
  \bibfield  {author} {\bibinfo {author} {\bibfnamefont {M.~A.}\ \bibnamefont {Castellanos-Beltran}}, \bibinfo {author} {\bibfnamefont {D.~I.}\ \bibnamefont {Olaya}}, \bibinfo {author} {\bibfnamefont {A.~J.}\ \bibnamefont {Sirois}}, \bibinfo {author} {\bibfnamefont {P.~D.}\ \bibnamefont {Dresselhaus}}, \bibinfo {author} {\bibfnamefont {S.~P.}\ \bibnamefont {Benz}}, \ and\ \bibinfo {author} {\bibfnamefont {P.~F.}\ \bibnamefont {Hopkins}},\ }\bibfield  {title} {\enquote {\bibinfo {title} {Stacked josephson junctions as inductors for single flux quantum circuits},}\ }\href {\doibase 10.1109/tasc.2019.2898406} {\bibfield  {journal} {\bibinfo  {journal} {IEEE Transactions on Applied Superconductivity}\ }\textbf {\bibinfo {volume} {29}},\ \bibinfo {pages} {1300705} (\bibinfo {year} {2019})}\BibitemShut {NoStop}%
\bibitem [{\citenamefont {Tolpygo}\ \emph {et~al.}(2023)\citenamefont {Tolpygo}, \citenamefont {Mallek}, \citenamefont {Bolkhovsky}, \citenamefont {Rastogi}, \citenamefont {Golden}, \citenamefont {Weir}, \citenamefont {Johnson},\ and\ \citenamefont {Gouker}}]{Tolpygo2023}%
  \BibitemOpen
  \bibfield  {author} {\bibinfo {author} {\bibfnamefont {S.~K.}\ \bibnamefont {Tolpygo}}, \bibinfo {author} {\bibfnamefont {J.~L.}\ \bibnamefont {Mallek}}, \bibinfo {author} {\bibfnamefont {V.}~\bibnamefont {Bolkhovsky}}, \bibinfo {author} {\bibfnamefont {R.}~\bibnamefont {Rastogi}}, \bibinfo {author} {\bibfnamefont {E.~B.}\ \bibnamefont {Golden}}, \bibinfo {author} {\bibfnamefont {T.~J.}\ \bibnamefont {Weir}}, \bibinfo {author} {\bibfnamefont {L.~M.}\ \bibnamefont {Johnson}}, \ and\ \bibinfo {author} {\bibfnamefont {M.~A.}\ \bibnamefont {Gouker}},\ }\bibfield  {title} {\enquote {\bibinfo {title} {Progress toward superconductor electronics fabrication process with planarized nbn and nbn/nb layers},}\ }\href {\doibase 10.1109/tasc.2023.3246430} {\bibfield  {journal} {\bibinfo  {journal} {IEEE Transactions on Applied Superconductivity}\ }\textbf {\bibinfo {volume} {33}},\ \bibinfo {pages} {1–12} (\bibinfo {year} {2023})}\BibitemShut {NoStop}%
\bibitem [{\citenamefont {Pechenezhskiy}\ \emph {et~al.}(2020)\citenamefont {Pechenezhskiy}, \citenamefont {Mencia}, \citenamefont {Nguyen}, \citenamefont {Lin},\ and\ \citenamefont {Manucharyan}}]{Pechenezhskiy2020}%
  \BibitemOpen
  \bibfield  {author} {\bibinfo {author} {\bibfnamefont {I.~V.}\ \bibnamefont {Pechenezhskiy}}, \bibinfo {author} {\bibfnamefont {R.~A.}\ \bibnamefont {Mencia}}, \bibinfo {author} {\bibfnamefont {L.~B.}\ \bibnamefont {Nguyen}}, \bibinfo {author} {\bibfnamefont {Y.-H.}\ \bibnamefont {Lin}}, \ and\ \bibinfo {author} {\bibfnamefont {V.~E.}\ \bibnamefont {Manucharyan}},\ }\bibfield  {title} {\enquote {\bibinfo {title} {The superconducting quasicharge qubit},}\ }\href {\doibase 10.1038/s41586-020-2687-9} {\bibfield  {journal} {\bibinfo  {journal} {Nature}\ }\textbf {\bibinfo {volume} {585}},\ \bibinfo {pages} {368–371} (\bibinfo {year} {2020})}\BibitemShut {NoStop}%
\bibitem [{\citenamefont {Bell}\ \emph {et~al.}(2012)\citenamefont {Bell}, \citenamefont {Sadovskyy}, \citenamefont {Ioffe}, \citenamefont {Kitaev},\ and\ \citenamefont {Gershenson}}]{Bell2012}%
  \BibitemOpen
  \bibfield  {author} {\bibinfo {author} {\bibfnamefont {M.~T.}\ \bibnamefont {Bell}}, \bibinfo {author} {\bibfnamefont {I.~A.}\ \bibnamefont {Sadovskyy}}, \bibinfo {author} {\bibfnamefont {L.~B.}\ \bibnamefont {Ioffe}}, \bibinfo {author} {\bibfnamefont {A.~Y.}\ \bibnamefont {Kitaev}}, \ and\ \bibinfo {author} {\bibfnamefont {M.~E.}\ \bibnamefont {Gershenson}},\ }\bibfield  {title} {\enquote {\bibinfo {title} {Quantum superinductor with tunable nonlinearity},}\ }\href {\doibase 10.1103/physrevlett.109.137003} {\bibfield  {journal} {\bibinfo  {journal} {Physical Review Letters}\ }\textbf {\bibinfo {volume} {109}} (\bibinfo {year} {2012}),\ 10.1103/physrevlett.109.137003}\BibitemShut {NoStop}%
\bibitem [{\citenamefont {Ustinov}\ \emph {et~al.}(1993)\citenamefont {Ustinov}, \citenamefont {Kohlstedt}, \citenamefont {Cirillo}, \citenamefont {Pedersen}, \citenamefont {Hallmanns},\ and\ \citenamefont {Heiden}}]{Ustinov1993}%
  \BibitemOpen
  \bibfield  {author} {\bibinfo {author} {\bibfnamefont {A.~V.}\ \bibnamefont {Ustinov}}, \bibinfo {author} {\bibfnamefont {H.}~\bibnamefont {Kohlstedt}}, \bibinfo {author} {\bibfnamefont {M.}~\bibnamefont {Cirillo}}, \bibinfo {author} {\bibfnamefont {N.~F.}\ \bibnamefont {Pedersen}}, \bibinfo {author} {\bibfnamefont {G.}~\bibnamefont {Hallmanns}}, \ and\ \bibinfo {author} {\bibfnamefont {C.}~\bibnamefont {Heiden}},\ }\bibfield  {title} {\enquote {\bibinfo {title} {Coupled fluxon modes in stacked nb/alox/nb long josephson junctions},}\ }\href {\doibase 10.1103/physrevb.48.10614} {\bibfield  {journal} {\bibinfo  {journal} {Physical Review B}\ }\textbf {\bibinfo {volume} {48}},\ \bibinfo {pages} {10614–10617} (\bibinfo {year} {1993})}\BibitemShut {NoStop}%
\bibitem [{\citenamefont {Sakai}\ \emph {et~al.}(1994)\citenamefont {Sakai}, \citenamefont {Ustinov}, \citenamefont {Kohlstedt}, \citenamefont {Petraglia},\ and\ \citenamefont {Pedersen}}]{Sakai1994}%
  \BibitemOpen
  \bibfield  {author} {\bibinfo {author} {\bibfnamefont {S.}~\bibnamefont {Sakai}}, \bibinfo {author} {\bibfnamefont {A.~V.}\ \bibnamefont {Ustinov}}, \bibinfo {author} {\bibfnamefont {H.}~\bibnamefont {Kohlstedt}}, \bibinfo {author} {\bibfnamefont {A.}~\bibnamefont {Petraglia}}, \ and\ \bibinfo {author} {\bibfnamefont {N.~F.}\ \bibnamefont {Pedersen}},\ }\bibfield  {title} {\enquote {\bibinfo {title} {Theory and experiment on electromagnetic-wave-propagation velocities in stacked superconducting tunnel structures},}\ }\href {\doibase 10.1103/physrevb.50.12905} {\bibfield  {journal} {\bibinfo  {journal} {Physical Review B}\ }\textbf {\bibinfo {volume} {50}},\ \bibinfo {pages} {12905–12914} (\bibinfo {year} {1994})}\BibitemShut {NoStop}%
\bibitem [{\citenamefont {Pedersen}\ and\ \citenamefont {Ustinov}(1995)}]{Pedersen1995}%
  \BibitemOpen
  \bibfield  {author} {\bibinfo {author} {\bibfnamefont {N.~F.}\ \bibnamefont {Pedersen}}\ and\ \bibinfo {author} {\bibfnamefont {A.~V.}\ \bibnamefont {Ustinov}},\ }\bibfield  {title} {\enquote {\bibinfo {title} {Fluxons in josephson transmission lines: new developments},}\ }\href {\doibase 10.1088/0953-2048/8/6/001} {\bibfield  {journal} {\bibinfo  {journal} {Superconductor Science and Technology}\ }\textbf {\bibinfo {volume} {8}},\ \bibinfo {pages} {389–401} (\bibinfo {year} {1995})}\BibitemShut {NoStop}%
\bibitem [{\citenamefont {Ustinov}\ and\ \citenamefont {Kohlstedt}(1996)}]{Ustinov1996}%
  \BibitemOpen
  \bibfield  {author} {\bibinfo {author} {\bibfnamefont {A.~V.}\ \bibnamefont {Ustinov}}\ and\ \bibinfo {author} {\bibfnamefont {H.}~\bibnamefont {Kohlstedt}},\ }\bibfield  {title} {\enquote {\bibinfo {title} {Interlayer fluxon interaction in josephson stacks},}\ }\href {\doibase 10.1103/physrevb.54.6111} {\bibfield  {journal} {\bibinfo  {journal} {Physical Review B}\ }\textbf {\bibinfo {volume} {54}},\ \bibinfo {pages} {6111–6114} (\bibinfo {year} {1996})}\BibitemShut {NoStop}%
\bibitem [{\citenamefont {Shitov}\ \emph {et~al.}(1996)\citenamefont {Shitov}, \citenamefont {Ustinov}, \citenamefont {Iosad},\ and\ \citenamefont {Kohlstedt}}]{Shitov1996}%
  \BibitemOpen
  \bibfield  {author} {\bibinfo {author} {\bibfnamefont {S.~V.}\ \bibnamefont {Shitov}}, \bibinfo {author} {\bibfnamefont {A.~V.}\ \bibnamefont {Ustinov}}, \bibinfo {author} {\bibfnamefont {N.}~\bibnamefont {Iosad}}, \ and\ \bibinfo {author} {\bibfnamefont {H.}~\bibnamefont {Kohlstedt}},\ }\bibfield  {title} {\enquote {\bibinfo {title} {On-chip radiation detection from stacked josephson flux-flow oscillators},}\ }\href {\doibase 10.1063/1.363734} {\bibfield  {journal} {\bibinfo  {journal} {Journal of Applied Physics}\ }\textbf {\bibinfo {volume} {80}},\ \bibinfo {pages} {7134–7137} (\bibinfo {year} {1996})}\BibitemShut {NoStop}%
\bibitem [{\citenamefont {Hutter}\ \emph {et~al.}(2011)\citenamefont {Hutter}, \citenamefont {Tholén}, \citenamefont {Stannigel}, \citenamefont {Lidmar},\ and\ \citenamefont {Haviland}}]{Hutter2011}%
  \BibitemOpen
  \bibfield  {author} {\bibinfo {author} {\bibfnamefont {C.}~\bibnamefont {Hutter}}, \bibinfo {author} {\bibfnamefont {E.~A.}\ \bibnamefont {Tholén}}, \bibinfo {author} {\bibfnamefont {K.}~\bibnamefont {Stannigel}}, \bibinfo {author} {\bibfnamefont {J.}~\bibnamefont {Lidmar}}, \ and\ \bibinfo {author} {\bibfnamefont {D.~B.}\ \bibnamefont {Haviland}},\ }\bibfield  {title} {\enquote {\bibinfo {title} {Josephson junction transmission lines as tunable artificial crystals},}\ }\href {\doibase 10.1103/physrevb.83.014511} {\bibfield  {journal} {\bibinfo  {journal} {Physical Review B}\ }\textbf {\bibinfo {volume} {83}} (\bibinfo {year} {2011}),\ 10.1103/physrevb.83.014511}\BibitemShut {NoStop}%
\bibitem [{\citenamefont {Masluk}\ \emph {et~al.}(2012)\citenamefont {Masluk}, \citenamefont {Pop}, \citenamefont {Kamal}, \citenamefont {Minev},\ and\ \citenamefont {Devoret}}]{Masluk2012}%
  \BibitemOpen
  \bibfield  {author} {\bibinfo {author} {\bibfnamefont {N.~A.}\ \bibnamefont {Masluk}}, \bibinfo {author} {\bibfnamefont {I.~M.}\ \bibnamefont {Pop}}, \bibinfo {author} {\bibfnamefont {A.}~\bibnamefont {Kamal}}, \bibinfo {author} {\bibfnamefont {Z.~K.}\ \bibnamefont {Minev}}, \ and\ \bibinfo {author} {\bibfnamefont {M.~H.}\ \bibnamefont {Devoret}},\ }\bibfield  {title} {\enquote {\bibinfo {title} {Microwave characterization of josephson junction arrays: Implementing a low loss superinductance},}\ }\href {\doibase 10.1103/physrevlett.109.137002} {\bibfield  {journal} {\bibinfo  {journal} {Physical Review Letters}\ }\textbf {\bibinfo {volume} {109}} (\bibinfo {year} {2012}),\ 10.1103/physrevlett.109.137002}\BibitemShut {NoStop}%
\bibitem [{\citenamefont {Maleeva}\ \emph {et~al.}(2018)\citenamefont {Maleeva}, \citenamefont {Gr\"{u}nhaupt}, \citenamefont {Klein}, \citenamefont {Levy-Bertrand}, \citenamefont {Dupre}, \citenamefont {Calvo}, \citenamefont {Valenti}, \citenamefont {Winkel}, \citenamefont {Friedrich}, \citenamefont {Wernsdorfer}, \citenamefont {Ustinov}, \citenamefont {Rotzinger}, \citenamefont {Monfardini}, \citenamefont {Fistul},\ and\ \citenamefont {Pop}}]{Maleeva2018}%
  \BibitemOpen
  \bibfield  {author} {\bibinfo {author} {\bibfnamefont {N.}~\bibnamefont {Maleeva}}, \bibinfo {author} {\bibfnamefont {L.}~\bibnamefont {Gr\"{u}nhaupt}}, \bibinfo {author} {\bibfnamefont {T.}~\bibnamefont {Klein}}, \bibinfo {author} {\bibfnamefont {F.}~\bibnamefont {Levy-Bertrand}}, \bibinfo {author} {\bibfnamefont {O.}~\bibnamefont {Dupre}}, \bibinfo {author} {\bibfnamefont {M.}~\bibnamefont {Calvo}}, \bibinfo {author} {\bibfnamefont {F.}~\bibnamefont {Valenti}}, \bibinfo {author} {\bibfnamefont {P.}~\bibnamefont {Winkel}}, \bibinfo {author} {\bibfnamefont {F.}~\bibnamefont {Friedrich}}, \bibinfo {author} {\bibfnamefont {W.}~\bibnamefont {Wernsdorfer}}, \bibinfo {author} {\bibfnamefont {A.~V.}\ \bibnamefont {Ustinov}}, \bibinfo {author} {\bibfnamefont {H.}~\bibnamefont {Rotzinger}}, \bibinfo {author} {\bibfnamefont {A.}~\bibnamefont {Monfardini}}, \bibinfo {author} {\bibfnamefont {M.~V.}\ \bibnamefont {Fistul}}, \ and\ \bibinfo {author} {\bibfnamefont {I.~M.}\ \bibnamefont {Pop}},\ }\bibfield  {title}
  {\enquote {\bibinfo {title} {Circuit quantum electrodynamics of granular aluminum resonators},}\ }\href {\doibase 10.1038/s41467-018-06386-9} {\bibfield  {journal} {\bibinfo  {journal} {Nature Communications}\ }\textbf {\bibinfo {volume} {9}} (\bibinfo {year} {2018}),\ 10.1038/s41467-018-06386-9}\BibitemShut {NoStop}%
\bibitem [{\citenamefont {Inc.}(2023)}]{ansys_maxwell}%
  \BibitemOpen
  \bibfield  {author} {\bibinfo {author} {\bibfnamefont {A.}~\bibnamefont {Inc.}},\ }\href@noop {} {\enquote {\bibinfo {title} {Ansys maxwell},}\ }\bibinfo {howpublished} {\url{https://www.ansys.com/products/electronics/ansys-maxwell}} (\bibinfo {year} {2023}),\ \bibinfo {note} {accessed: 2023-12-20}\BibitemShut {NoStop}%
\bibitem [{\citenamefont {Haviland}, \citenamefont {Andersson},\ and\ \citenamefont {Ågren}(2000)}]{Haviland2000}%
  \BibitemOpen
  \bibfield  {author} {\bibinfo {author} {\bibfnamefont {D.~B.}\ \bibnamefont {Haviland}}, \bibinfo {author} {\bibfnamefont {K.}~\bibnamefont {Andersson}}, \ and\ \bibinfo {author} {\bibfnamefont {P.}~\bibnamefont {Ågren}},\ }\bibfield  {title} {\enquote {\bibinfo {title} {Superconducting and insulating behavior in one-dimensional josephson junction arrays},}\ }\href {\doibase 10.1023/a:1004603814529} {\bibfield  {journal} {\bibinfo  {journal} {Journal of Low Temperature Physics}\ }\textbf {\bibinfo {volume} {118}},\ \bibinfo {pages} {733–749} (\bibinfo {year} {2000})}\BibitemShut {NoStop}%
\bibitem [{\citenamefont {Ó~Peatáin}\ \emph {et~al.}(2023)\citenamefont {Ó~Peatáin}, \citenamefont {Dixon}, \citenamefont {Meeson}, \citenamefont {Williams}, \citenamefont {Kafanov},\ and\ \citenamefont {Pashkin}}]{Peatin2023}%
  \BibitemOpen
  \bibfield  {author} {\bibinfo {author} {\bibfnamefont {S.}~\bibnamefont {Ó~Peatáin}}, \bibinfo {author} {\bibfnamefont {T.}~\bibnamefont {Dixon}}, \bibinfo {author} {\bibfnamefont {P.~J.}\ \bibnamefont {Meeson}}, \bibinfo {author} {\bibfnamefont {J.~M.}\ \bibnamefont {Williams}}, \bibinfo {author} {\bibfnamefont {S.}~\bibnamefont {Kafanov}}, \ and\ \bibinfo {author} {\bibfnamefont {Y.~A.}\ \bibnamefont {Pashkin}},\ }\bibfield  {title} {\enquote {\bibinfo {title} {Simulating the effects of fabrication tolerance on the performance of josephson travelling wave parametric amplifiers},}\ }\href {\doibase 10.1088/1361-6668/acba4e} {\bibfield  {journal} {\bibinfo  {journal} {Superconductor Science and Technology}\ }\textbf {\bibinfo {volume} {36}},\ \bibinfo {pages} {045017} (\bibinfo {year} {2023})}\BibitemShut {NoStop}%
\bibitem [{\citenamefont {Feldman}\ and\ \citenamefont {Rogachev}(2024)}]{Feldman2024}%
  \BibitemOpen
  \bibfield  {author} {\bibinfo {author} {\bibfnamefont {S.}~\bibnamefont {Feldman}}\ and\ \bibinfo {author} {\bibfnamefont {A.}~\bibnamefont {Rogachev}},\ }\href {\doibase 10.48550/ARXIV.2411.06492} {\enquote {\bibinfo {title} {Quantum phase transition in small-size 1d and 2d josephson junction arrays: analysis of the experiments within the interacting plasmons picture},}\ } (\bibinfo {year} {2024})\BibitemShut {NoStop}%
\bibitem [{\citenamefont {Weißl}\ \emph {et~al.}(2015)\citenamefont {Weißl}, \citenamefont {K\"{u}ng}, \citenamefont {Dumur}, \citenamefont {Feofanov}, \citenamefont {Matei}, \citenamefont {Naud}, \citenamefont {Buisson}, \citenamefont {Hekking},\ and\ \citenamefont {Guichard}}]{Weil2015}%
  \BibitemOpen
  \bibfield  {author} {\bibinfo {author} {\bibfnamefont {T.}~\bibnamefont {Weißl}}, \bibinfo {author} {\bibfnamefont {B.}~\bibnamefont {K\"{u}ng}}, \bibinfo {author} {\bibfnamefont {E.}~\bibnamefont {Dumur}}, \bibinfo {author} {\bibfnamefont {A.~K.}\ \bibnamefont {Feofanov}}, \bibinfo {author} {\bibfnamefont {I.}~\bibnamefont {Matei}}, \bibinfo {author} {\bibfnamefont {C.}~\bibnamefont {Naud}}, \bibinfo {author} {\bibfnamefont {O.}~\bibnamefont {Buisson}}, \bibinfo {author} {\bibfnamefont {F.~W.~J.}\ \bibnamefont {Hekking}}, \ and\ \bibinfo {author} {\bibfnamefont {W.}~\bibnamefont {Guichard}},\ }\bibfield  {title} {\enquote {\bibinfo {title} {Kerr coefficients of plasma resonances in josephson junction chains},}\ }\href {\doibase 10.1103/physrevb.92.104508} {\bibfield  {journal} {\bibinfo  {journal} {Physical Review B}\ }\textbf {\bibinfo {volume} {92}} (\bibinfo {year} {2015}),\ 10.1103/physrevb.92.104508}\BibitemShut {NoStop}%
\bibitem [{\citenamefont {Ambegaokar}\ and\ \citenamefont {Baratoff}(1963)}]{Ambegaokar1963}%
  \BibitemOpen
  \bibfield  {author} {\bibinfo {author} {\bibfnamefont {V.}~\bibnamefont {Ambegaokar}}\ and\ \bibinfo {author} {\bibfnamefont {A.}~\bibnamefont {Baratoff}},\ }\bibfield  {title} {\enquote {\bibinfo {title} {Tunneling between superconductors},}\ }\href {\doibase 10.1103/physrevlett.10.486} {\bibfield  {journal} {\bibinfo  {journal} {Physical Review Letters}\ }\textbf {\bibinfo {volume} {10}},\ \bibinfo {pages} {486–489} (\bibinfo {year} {1963})}\BibitemShut {NoStop}%
\bibitem [{\citenamefont {Lotkhov}\ \emph {et~al.}(2006)\citenamefont {Lotkhov}, \citenamefont {Tolkacheva}, \citenamefont {Balashov}, \citenamefont {Khabipov}, \citenamefont {Buchholz},\ and\ \citenamefont {Zorin}}]{Lotkhov2006}%
  \BibitemOpen
  \bibfield  {author} {\bibinfo {author} {\bibfnamefont {S.~V.}\ \bibnamefont {Lotkhov}}, \bibinfo {author} {\bibfnamefont {E.~M.}\ \bibnamefont {Tolkacheva}}, \bibinfo {author} {\bibfnamefont {D.~V.}\ \bibnamefont {Balashov}}, \bibinfo {author} {\bibfnamefont {M.~I.}\ \bibnamefont {Khabipov}}, \bibinfo {author} {\bibfnamefont {F.-I.}\ \bibnamefont {Buchholz}}, \ and\ \bibinfo {author} {\bibfnamefont {A.~B.}\ \bibnamefont {Zorin}},\ }\bibfield  {title} {\enquote {\bibinfo {title} {Low hysteretic behavior of al/alox/al josephson junctions},}\ }\href {\doibase 10.1063/1.2357915} {\bibfield  {journal} {\bibinfo  {journal} {Applied Physics Letters}\ }\textbf {\bibinfo {volume} {89}} (\bibinfo {year} {2006}),\ 10.1063/1.2357915}\BibitemShut {NoStop}%
\bibitem [{\citenamefont {Basko}\ and\ \citenamefont {Hekking}(2013)}]{Basko2013}%
  \BibitemOpen
  \bibfield  {author} {\bibinfo {author} {\bibfnamefont {D.~M.}\ \bibnamefont {Basko}}\ and\ \bibinfo {author} {\bibfnamefont {F.~W.~J.}\ \bibnamefont {Hekking}},\ }\bibfield  {title} {\enquote {\bibinfo {title} {Disordered josephson junction chains: Anderson localization of normal modes and impedance fluctuations},}\ }\href {\doibase 10.1103/physrevb.88.094507} {\bibfield  {journal} {\bibinfo  {journal} {Physical Review B}\ }\textbf {\bibinfo {volume} {88}} (\bibinfo {year} {2013}),\ 10.1103/physrevb.88.094507}\BibitemShut {NoStop}%
\bibitem [{\citenamefont {Nguyen}\ and\ \citenamefont {Basko}(2017)}]{Nguyen2017}%
  \BibitemOpen
  \bibfield  {author} {\bibinfo {author} {\bibfnamefont {D.~V.}\ \bibnamefont {Nguyen}}\ and\ \bibinfo {author} {\bibfnamefont {D.~M.}\ \bibnamefont {Basko}},\ }\bibfield  {title} {\enquote {\bibinfo {title} {Inhomogeneous josephson junction chains: a superconducting meta-material for superinductance optimization},}\ }\href {\doibase 10.1140/epjst/e2016-60278-4} {\bibfield  {journal} {\bibinfo  {journal} {The European Physical Journal Special Topics}\ }\textbf {\bibinfo {volume} {226}},\ \bibinfo {pages} {1499–1514} (\bibinfo {year} {2017})}\BibitemShut {NoStop}%
\bibitem [{\citenamefont {Kleinsasser}, \citenamefont {Miller},\ and\ \citenamefont {Mallison}(1995)}]{Kleinsasser1995}%
  \BibitemOpen
  \bibfield  {author} {\bibinfo {author} {\bibfnamefont {A.}~\bibnamefont {Kleinsasser}}, \bibinfo {author} {\bibfnamefont {R.}~\bibnamefont {Miller}}, \ and\ \bibinfo {author} {\bibfnamefont {W.}~\bibnamefont {Mallison}},\ }\bibfield  {title} {\enquote {\bibinfo {title} {Dependence of critical current density on oxygen exposure in nb-alo/sub x/-nb tunnel junctions},}\ }\href {\doibase 10.1109/77.384565} {\bibfield  {journal} {\bibinfo  {journal} {IEEE Transactions on Appiled Superconductivity}\ }\textbf {\bibinfo {volume} {5}},\ \bibinfo {pages} {26–30} (\bibinfo {year} {1995})}\BibitemShut {NoStop}%
\end{thebibliography}%
\end{document}